\documentclass[10pt,journal,compsoc]{IEEEtran}
\usepackage{csquotes}
\usepackage{booktabs}

\usepackage{graphicx}
\usepackage{subcaption}
\usepackage{todonotes}
\usepackage{pdfpages}

\ifCLASSOPTIONcompsoc
  \usepackage[nocompress]{cite}
\else
  \usepackage{cite}
\fi

\usepackage[numbers]{natbib}

\ifCLASSINFOpdf
\else
\fi
\usepackage{amsmath,amsfonts,bm,bbm}
\newcommand\MYhyperrefoptions{bookmarks=true,bookmarksnumbered=true,
pdfpagemode={UseOutlines},plainpages=false,pdfpagelabels=true,
colorlinks=true,linkcolor={black},citecolor={black},urlcolor={black},
pdftitle={Bare Demo of IEEEtran.cls for Computer Society Journals},%<!CHANGE!
pdfsubject={Typesetting},%<!CHANGE!
pdfauthor={Michael D. Shell},%<!CHANGE!
pdfkeywords={Computer Society, IEEEtran, journal, LaTeX, paper,
             template}}%<^!CHANGE!
\ifCLASSINFOpdf
\usepackage[\MYhyperrefoptions,pdftex]{hyperref}
\else
\usepackage[\MYhyperrefoptions,breaklinks=true,dvips]{hyperref}
\usepackage{breakurl}
\fi
\begin{document}
\title{Perceived Conversation Quality in\\ Spontaneous Interactions}

\author{Chirag~Raman$^{\star}$~\IEEEmembership{Non-Member,~IEEE,} Navin~Raj Prabhu$^{\star}$~\IEEEmembership{Member,~IEEE,}
Hayley~Hung~\IEEEmembership{Member,~IEEE}% <-this % stops a space
\IEEEcompsocitemizethanks{
\IEEEcompsocthanksitem Chirag~Raman and Hayley~Hung are with the Intelligent Systems, Delft University of Technology EEMCS, 225112 Delft, Zuid-Holland, Netherlands, 2628XE (e-mail: c.a.raman@tudelft.nl, h.hung@tudelft.nl) \\
\IEEEcompsocthanksitem Navin~Raj Prabhu was with the Intelligent Systems, Delft University of Technology EEMCS, 225112 Delft, Zuid-Holland, Netherlands, 2628XE (e-mail: lr.navin@yahoo.nl).
\protect\\}
\thanks{$^{\star}$ Both authors contributed equally to this work.}
}

\markboth{Journal of \LaTeX\ Class Files,~Vol.~14, No.~8, August~2015}%
{Shell \MakeLowercase{\textit{et al.}}: Bare Advanced Demo of IEEEtran.cls for IEEE Computer Society Journals}



\IEEEtitleabstractindextext{%
\begin{abstract}
The quality of daily spontaneous conversations is of importance towards both our well-being as well as the development of interactive social agents. Prior research directly studying the quality of social conversations has operationalized it in narrow terms, associating greater quality to less small talk. Other works taking a broader perspective of interaction experience have indirectly studied quality through one of the several overlapping constructs such as rapport or engagement, in isolation. In this work we bridge this gap by proposing a holistic conceptualization of conversation quality, building upon the collaborative attributes of cooperative conversation floors. Taking a multilevel perspective of conversation, we develop and validate two instruments for perceived conversation quality (PCQ) at the individual and group levels. Specifically, we motivate capturing external raters' gestalt impressions of participant experiences from thin slices of behavior, and collect annotations of PCQ on the publicly available MatchNMingle dataset of in-the-wild mingling conversations. Finally, we present an analysis of behavioral features that are predictive of PCQ. We find that for the conversations in MatchNMingle, raters tend to associate smaller group sizes, equitable speaking turns with fewer interruptions, and time taken for synchronous bodily coordination with higher PCQ.
\end{abstract}

}

\maketitle

% is not selected <OR> if conference mode is selected - because compsoc
% conference papers position the abstract like regular (non-compsoc)
\IEEEdisplaynontitleabstractindextext

% \fi
\IEEEpeerreviewmaketitle

\ifCLASSOPTIONcompsoc
\IEEEraisesectionheading{\section{Introduction}\label{sec:introduction}}
\else
\section{Introduction}
\label{sec:introduction}
\fi

Picture a spontaneous interaction such as a daily social conversation at work or home. The quality of such conversations is of importance towards both our well-being as well as the development of interactive technologies that influence our daily lives. At an individual level, conversation quality is directly associated with our happiness and life satisfaction \cite{milekEavesdroppingHappinessRevisited, mehlEavesdroppingHappiness}. Furthermore, human judgement of conversation quality is a common measure for the evaluation of artificial conversation agents \cite{seeWhatMakesGood2019}. Despite its importance, little prior research has directly studied conversation quality or jointly considered the factors affecting its perception. 

\begin{figure}[t]
    \centering
    \includegraphics[width=0.8\linewidth]{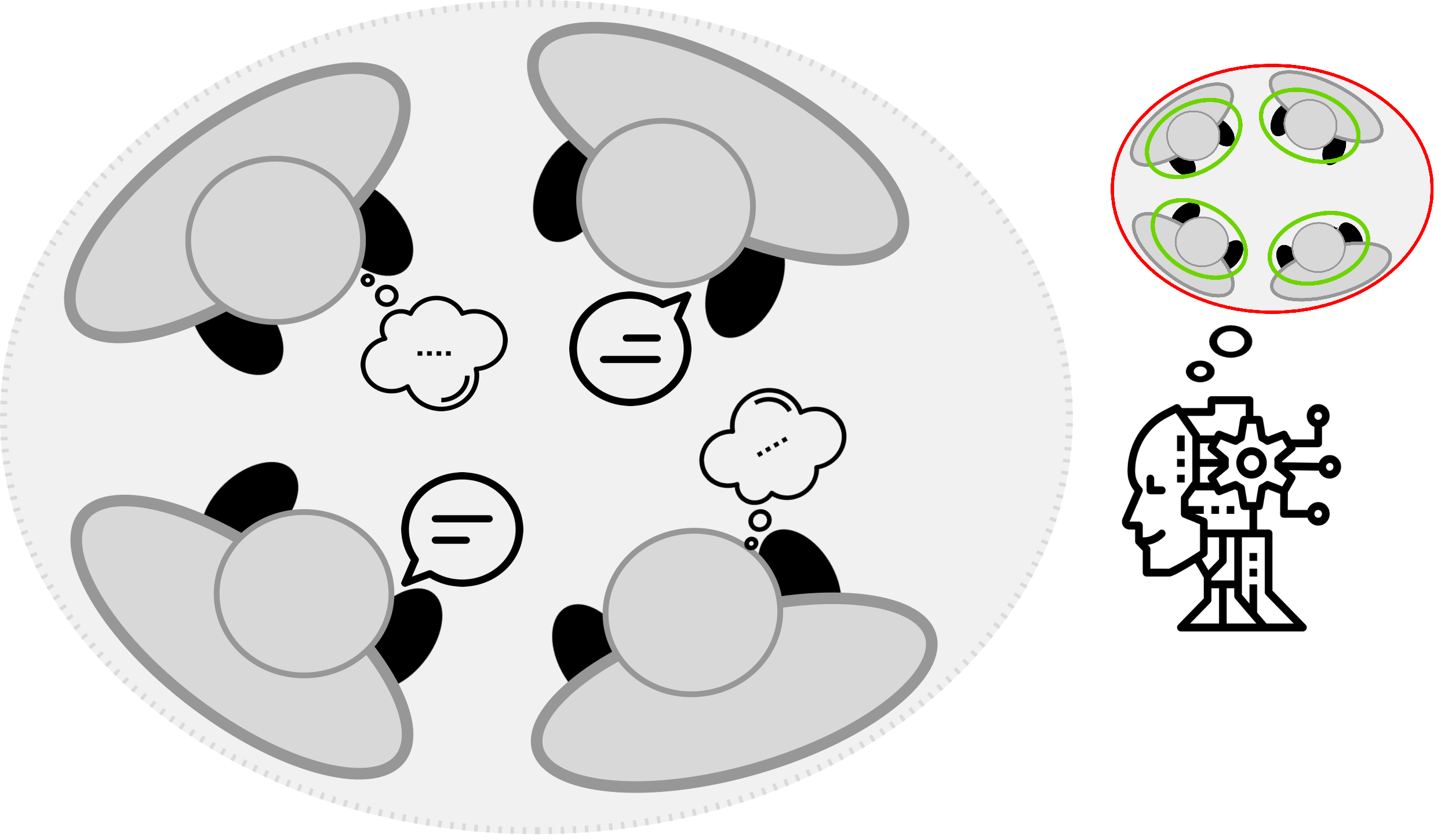}
    \caption{Conceptual illustration of individual experiences existing in the perception of interacting partners, and how an external perceived measure of individual-level (green) and group-level (red) experience is relevant for the development artificial interactive social agents.}
    \label{fig:intro}
    \vspace{-10pt}
\end{figure}

One challenge is that conversation quality is not directly measured, and needs to be inferred from observable verbal and non-verbal behavioral cues. This has led to some research viewing conversation quality in narrow terms, considering only isolated attributes of the conversation. For instance, \citet{milekEavesdroppingHappinessRevisited} and \citet{mehlEavesdroppingHappiness} consider greater conversation quality to correspond to less small talk and information exchange at more than a trivial level of depth. On the other hand, taking a broader view of conversation quality runs into another challenge: its potential intersection with several overlapping social concepts. These include rapport \cite{rapport_muller},  bonding \cite{Jaques2016}, interest-levels \cite{gatica_interest}, and involvement \cite{catha_involve} amongst others. When studied towards the development of interactive systems \cite{seeWhatMakesGood2019}, human judgement of quality has been evaluated simply as a measure of enjoyment. Moreover, the focus here has been on the quality of artificially generated dialogue acts, and only in dyadic conversations. So while such factors have been studied individually in isolation, a joint consideration of the multiple aspects of conversation quality largely remains a knowledge gap. In this work, we take the perspective that such a \textit{holistic} characterization of the quality of multiparty spontaneous interactions is an important objective in the development of socially intelligent systems. For instance, consider a social robot approaching a conversing group of people, as illustrated in \figurename~\ref{fig:intro}. Here, a perception of the group's experience of the conversation as a whole could aid the social agent in developing more nuanced policies of approach. Furthermore, an estimate of each individual's experience could then aid the agent in developing personalized adaptive strategies to smoothly conduct the subsequent interaction.

In addition to a holistic characterization, we specifically argue for a \textit{perceived} measure of conversation quality in this work, at both the individual and the group level.
This is in contrast to existing efforts for quantifying quality-related aspects of conversations, which have largely focused on self-reported measures after interactions \cite{Hagad_rapport, rapport_muller, Jaques2016, Cuperman2009}. While such measures attempt to estimate an individual's true experience in-situ, they also suffer several drawbacks including desirability bias \cite{northrup1997problem}, egoistic bias \cite{northrup1997problem, garcia1997science}, and recall bias and cognitive errors \cite{backdated_esm}. On the other hand, a perceived measure of experience quantifies how participants seem to be experiencing the interaction to an external third-party observer \cite{gatica_interest, Lindley2013, catha_involve}. While such a measure may not capture the true experience, it closely models how we conduct interactions based on imperfect estimates of our conversation partners' experiences, and is therefore also useful towards the development of machines with social intelligence.

Concretely, we make three contributions in this work. First, we introduce the novel measure of \textit{Perceived Conversation Quality} (PCQ) towards quantifying social experience in spontaneous interactions, by jointly considering potentially overlapping related constructs. Second, we present an instrument for collecting annotations of PCQ at both the individual and the group level. We validate the instrument and collect annotations on the publicly available MatchNMingle dataset \cite{MatchNMingle_data}, and provide an analysis of the annotations. Third, we present insights into the behavioral features that predict PCQ through confirmatory statistical analysis and empirical data-driven analysis.

% To the best of our knowledge, there is no existing work in the literature which has attempted to define and quantify the overall perceived quality of spontaneous interactions from a holistic viewpoint.

%  \CR{[TODO: Make stronger, predictive modeling?]}

The rest of this work is organized as follows. In Section~\ref{chap:lit-review}, we review relevant literature. In Section~\ref{ConvQ-Section}, we conceptualize and define PCQ. In Section~\ref{Section:Annotations}, we specify our procedure for collecting and validating annotations of PCQ for interactions in-the-wild. In Section~\ref{Sec:ModelingConvQ}, we present details behavioral feature extraction and experimental setup, followed by the results in Section~\ref{Sec:Exp}. Finally, we conclude with a discussion of our findings in Section~\ref{sec:conclusion}. Our preliminary work on this topic was presented in \cite{raj2020defining}, which described the proposed instrument and analysis of annotations. The experiments we present in this manuscript (Section~\ref{Sec:ModelingConvQ} onward) are completely new. Moreover, this manuscript is a complete rewrite; compared to our prior publication the manuscript now includes a clearer (i) overall presentation and motivation, (ii) organization of related literature, and (iii) description of the process of conceptualizing, validating, and analyzing PCQ.   

% in Section~\ref{Sec:ModelingConvQ} followed by experiments and results in Section~\ref{Sec:Exp}. We discuss several key findings, limitations, and future avenues in Section~\ref{Discussion-Section}, finally concluding with Section~\ref{Conclusion-Section}.

% \todo[inline]{To Edit Related Work, wrt new content}

\section{Related work}\label{chap:lit-review}
Spontaneous interactions are considered to be non task-directed, unconstrained, and typically occurring in natural situations \cite{reitter2006priming, Oertel2011, wyatt_spont}. In such a dynamic conversation setting, several constructs emerge. These include descriptors of interpersonal relationships amongst participants (e.g. rapport \cite{rapport_muller} and bonding \cite{Jaques2016}), or those which capture qualitative attributes of the interaction (e.g. involvement \cite{antil1984conceptualization,catha_involve}, engagement \cite{hsiao_tt_engage}, and interest-levels \cite{gatica_interest}).

\subsection{Rapport and Bonding} Rapport and bonding have been widely studied as a pairwise phenomena using self-reported measures \cite{Hagad_rapport, rapport_muller, Jaques2016}. \citet{rapport_muller} define rapport as ``the close and harmonious relationship in which interaction partners are `in sync' with each other''. The authors used a self-reported questionnaire adapted from \citet{bernieri1996dyad} to measure rapport for every pair of individuals within small interaction groups. Another related social concept is bonding, which measures positive personal attachment including ``mutual trust, acceptance, and confidence'' amongst interacting pairs \cite{horvath1989development}. Based on this definition, \citet{Jaques2016} studied bonding in human-agent interactions, using the bonding subscale of the \textit{Working Alliance Inventory} (B-WAI) \cite{horvath1989development}.

\subsection{Involvement, Engagement, and Interest-Levels}
\citet{antil1984conceptualization} defines involvement as ``the level of perceived personal importance and/or interest evoked by a stimulus (or stimuli) within a specific situation''. Following \citeauthor{antil1984conceptualization}'s view of involvement as a non-binary variable, \citet{catha_involve} developed a $10$-level annotation scheme for joint involvement of a group based on intuitive, listener-independent impressions of prosody and body and face movement. \citet{oertel2013gaze} proposed a gaze-based method to relate group involvement to individual engagement in multiparty dialogue. Several researchers have conceptualized group cohesion to study its influence on task performance \cite{CohesionDef}, in settings such as meetings \cite{Behavior2010, Nanninga2017} and long-term crew missions \cite{YanxiaZhang2018_ITeam, Zhang2018_TeamSense}.
\citet{gatica_interest} define group interest-levels as, ``the perceived degree of interest or involvement of the majority of the group''. The authors provided perceived annotations for interest-levels using audio-visual recordings of interactions, on a discrete $5$-point scale. To this end, the external annotators were instructed to attend to interest-indicating activities such as note-taking, focused gaze, and avid participation in discussion. Note that these constructs have all been defined and studied in task-directed settings.

% From the above discussed literature, we see that research works tend to quantify constructs either by relying on self-reported measures or externally perceived annotations. Self-report measures have many advantages as they are the closest to the true experience, however they also come with several disadvantages. For example, they suffer from desirability bias \cite{northrup1997problem} and general behaviour of self-reporting interlocutors \cite{garcia1997science}. On the other hand, perceived measures are free from issues such as egoistic bias \cite{northrup1997problem, garcia1997science}, recall bias \cite{backdated_esm} and cognitive errors \cite{backdated_esm}. The characteristics of perceived measures, as discussed above, make them more suitable towards the development of social robots in dynamic spontaneous interactions. 
% several keys issues faced by self-reported measures, mainly the
% Also, in mingling scenarios where short bursts of spontaneous interaction occur, collecting interlocutor self-reports might be cumbersome as they are prone to intrusion and recall bias. At the same time, perceived measures are only an approximation of an individual's experience.

% In cases of series of short bursts of spontaneous interaction, subjects may tend to also forget longitudinal details and require an Experience Sampling Method (ESM) based data collection.

\subsection{General Measures of Interaction Experience}
In contrast to efforts focusing on specific social concepts, some recent approaches have proposed more general measures of experience in conversations.  
\citet{Cuperman2009} introduced the \textit{Perception of Interaction} (POI) questionnaire as part of a study to examine the effects of gender and personality traits on participant behaviors in dyadic interactions. The questionnaire collected self-reported measures of a participant's perception of their interaction experience. These aspects included the perceived quality of the interaction, the degree of rapport they felt they had with the other person, and the degree to which they liked the other person. This measure of interactions has been adapted by other works to study bonding \cite{Jaques2016} and interaction experience \cite{Cerekovic2014}. \citet{Lindley2013} follow the rationale that experience itself is difficult to quantify, but since it is entwined with social interaction, we might characterize experience by measuring aspects of conversation that are related to it. 
They studied several behavioral process measures and developed the \textit{Thin-Slice Enjoyment Scale} (TES): a measure of empathised enjoyment in social conversations from ratings of thin slices of behavior by na\"ive judges. In their factor analysis, the authors found that the judges viewed enjoyment and conversation fluency as being related. However, the POI was developed for self-reported measures, and neither work considered spontaneous interaction settings: \citet{Cuperman2009} considered scripted dyadic interactions with confederates, while \citet{Lindley2013} developed the TES within the particular task-directed context of photo sharing.

\section{Perceived Conversation Quality} \label{ConvQ-Section}

\subsection{Initial Conceptualization}
The primary influences for our conceptualization of PCQ are the works of \citet{edelsky-floor}, \citet{Lindley2013}, and \citet{Cuperman2009}. Specifically, from these and other works we motivate the rationale behind our choices of (i) focusing on the cooperative aspects of conversation towards conceptualizing PCQ, and (ii) rating thin slices of behavior to capture the gestalt impressions raters have of the continually unfolding conversation. 

In an analysis of social interactions in a series of meetings, \citet{edelsky-floor} observed two contrasting styles of conversation, termed \textit{cooperative floors} and \textit{exclusive floors}. Cooperative floors are characterized by collaborative stretches of ``free-for-all'' conversation accompanied by a feeling of participants being ``on the same wavelength'' \cite[p.~391]{edelsky-floor}. (In contrast, the exclusive floor is owned by a single person with turns rarely overlapping.) This notion of the cooperative floor captures the sense of engagement associated with positive experiences, and has been since linked with informal social interactions \cite{coates1989gossip, dunne1994simultaneous, tannen2005conversational} and enjoyment \cite{monkTelephoneConferencesFun}. As such, we observe that \citeauthor{edelsky-floor}'s notion of ``on the same wavelength'' strongly resonates with the POI questionnaire's focus on how interaction partners relate to each other \cite{Cuperman2009}.  Subsequent researchers have also derived qualitative measures of conversation based on the ``free-for-all'' aspects of \citeauthor{edelsky-floor}'s description. These include conversational equality and freedom \cite{Lindley2013} (or interactivity \cite{carletta1998placement}), and fluency through the occurrence of frequent turns \cite{Lindley2013, daly1998some}.

\citet{ambady1992thin} propose that thin slice judgements of behavior can be usefully made so long as the variables in question are observable and there is an affective or interpersonal component. They suggest that this is because such inferences are made through subconscious decoding of expressive behavior, with judgemental accuracy being strongly linked to ``gestalt, molar impressions based on nonverbal behavior'' \cite[p.~439]{ambady1993half}. This result supports previous research showing that molar impressions, although vaguer and fuzzier, generally yield more useful information than the coding of specific behaviors without accounting for overall context. Researchers often encourage the formation of this gestalt impression by intentionally reducing information presented to raters, e.g. removing speech content while retaining tone of voice or extinguishing facial expressions \cite{bernieri1994interactional}. In contrast, obtaining judgements of gestalt impressions is a natural fit for spontaneous interaction settings where recording speech or ego-centric perspectives is often not possible to preserve privacy \cite{MatchNMingle_data, raman2021social}.

\subsection{Pilot Qualitative Interviews with Na\"ive Judges}
\begin{figure}[!t]
    \centering
    \includegraphics[width=\columnwidth]{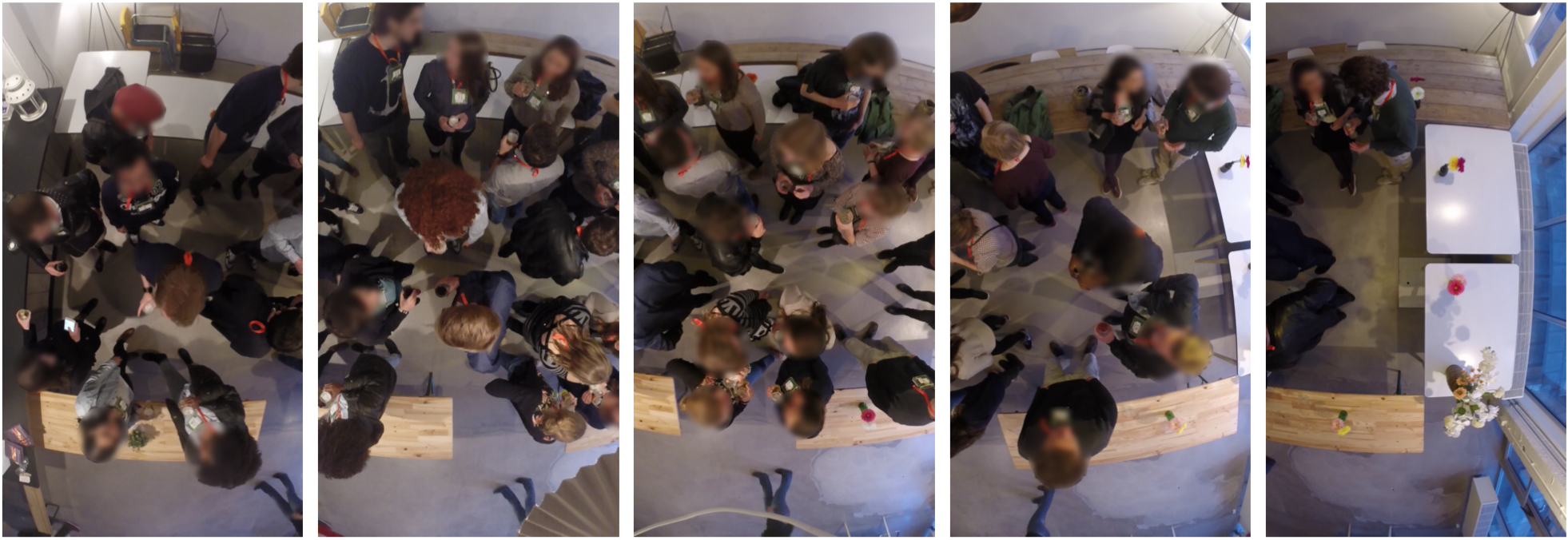}
    \caption{A snapshot from the MatchNMingle dataset \cite{cabrera2018matchnmingle}.}
    \label{fig:mnm}
\end{figure}

We conducted pilot qualitative interviews with three na\"ive judges to verify if our initial conceptualization matched the lay interpretation of PCQ. All judges were students enrolled in technical Masters programs at the authors' university. The judges were shown unaltered recordings from the publicly available MatchNMingle (MnM) dataset \cite{cabrera2018matchnmingle}, and asked what they thought of the conversations in the scene. \figurename~\ref{fig:mnm} illustrates a snapshot of a scene from MnM. To obtain unbiased impressions, we didn't specify our focus on conversation quality, nor our conceptualization of it. All judges (i) described a continually evolving perception of participant experiences over the conversation lifetime, aligning with our choice of rating thin slices of behavior rather than a single rating for the entire conversation; (ii) described perception of individual experiences as well as the group as a whole, aligning with our choice of measuring PCQ at the individual- and group- levels separately; and (iii) identified the attributes of equal opportunity for speaking, smoothness of interaction, and interpersonal relationships that strongly resonates with the prior work that serves as our primary influences \cite{edelsky-floor, Cuperman2009, Lindley2013}.  

\subsection{Definition and Constituents} \label{ConvQ-Section-Constituents}

Following our initial conceptualization and pilot interviews, we formalize PCQ of a spontaneous interaction as 
\begin{quote}
 \textit{the degree to which participants in the spontaneous interaction appear to be on the same wavelength and maintain an equal opportunity floor, as perceived by an external observer}.
\end{quote}

Further, in the following subsections we present three constituents of PCQ that categorize the multiple social concepts associated with this definition.

\subsubsection{Interpersonal Relationships}
This constituent describes the degree of association between participants or the notion of being in-sync with one's interaction partners, using constructs such as rapport \cite{rapport_muller} and bonding \cite{Jaques2016}. More specifically, the constituent measures the degree to which an individual was accepted and respected by other individuals in the group or the degree to which the other individuals were paying attention to the individual. Increased bonding and rapport amongst interacting partners is widely acknowledged to result in improved collaboration, and improved interpersonal outcomes, thereby having a key influence on the PCQ. 

\subsubsection{Nature of Interaction}
This constituent describes the degree to which the interaction was smooth and relaxed or forced and awkward. It captures the notion of whether the participants are having a positive and pleasant experience, drawing upon the quality of interaction aspects of the POI \cite{Cuperman2009}.

\subsubsection{Equal Opportunity}
This constituent captures the \textit{free-for-all} collaborative aspects of \citeauthor{edelsky-floor}'s description of cooperative floors \cite{edelsky-floor}. It describes the notion of equality of opportunity for participation shared amongst interacting partners, capturing the sense of cohesiveness and engagement in informal conversations. This includes factors such as conversation freedom \cite{Lai2018}, equality, and fluency \cite{Lindley2013} and an individual's opportunity to take the lead in the conversation \cite{Cuperman2009, Jaques2016}.

\begin{figure}[t]
\centering
\includegraphics[width=0.6\linewidth]{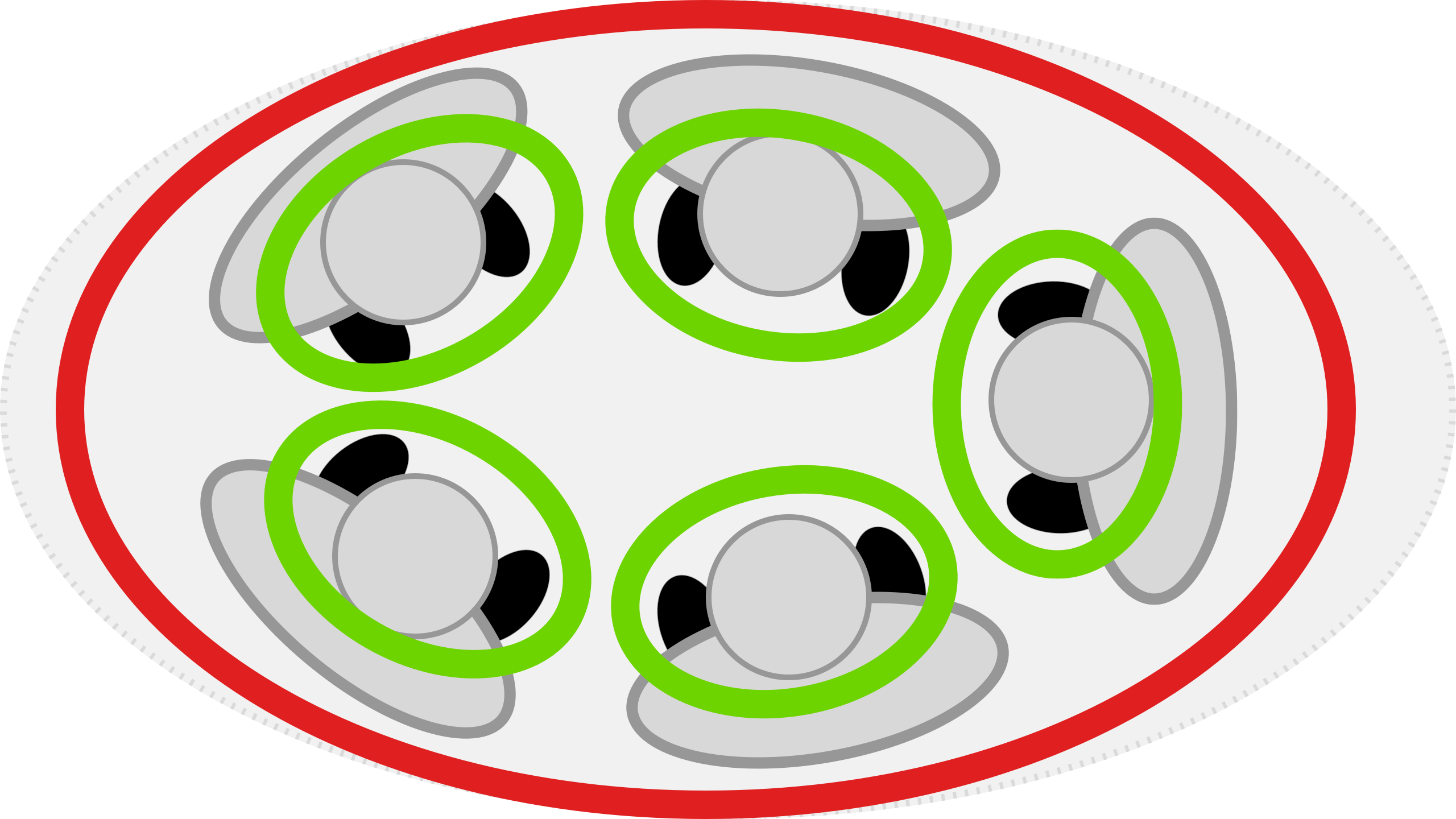}
% \captionsetup{justification=centering}
\caption{Illustrating the scope of observation to measure the group-level (red) and individual-level (green) PCQ.}
\label{fig:convq-manifest}
\end{figure}

\subsection{PCQ Questionnaires: A Multilevel Perspective} \label{ConvQ-Section-PercepForms}

We devise two independent questionnaires to measure PCQ at the individual and group levels. This follows our broader multilevel perspective \cite{Kozlowski2000} of social interactions where constructs can be conceptualized at different levels, such as the individual, dyadic, and group levels. While prior works have often considered constructs at a single level (e.g. \citet{rapport_muller} consider rapport as a dyadic pairwise construct), a multilevel perspective aligns better with our pilot judges' descriptions of attributes pertaining to individuals and groups as a whole. Moreover, some prior works on conversation group dynamics have indeed also taken a multilevel perspective: \citet{oertel2013gaze} distinguish overall group involvement from individual engagement, obtaining separate annotations at both levels. In the case of PCQ, our view is that an observer's perceptions of individual affect and behavior dynamically interact to contribute to an overall group-level perception. \figurename~\ref{fig:convq-manifest} illustrates the scope of observations towards measuring PCQ at each level.  

The individual level captures what the quality of the conversation appears to be to a particular individual. The focus is on how the individual seems to be relating to their partners and participating in the conversation. Consequently, every individual receives a rating. Note that this perspective doesn't consider the individual's behavior in \textit{isolation} by excluding the context of partner behaviors. Rather, the scope of consideration is restricted to what the individual seems to be experiencing. In contrast, the group level expands this scope of consideration to all interlocutors \textit{as a whole}, focusing on their collective experience, resulting in a single group-level rating.

Concretely, we devise the PCQ questionnaires by drawing upon elements of the POI scale \cite{Cuperman2009} and the TES \cite{Lindley2013}. However, since the POI was developed for self-reports rather than external perception, and neither was developed for spontaneous interaction settings, we adapt the specific items. First, all items were updated to address external observers and apply to group sizes beyond dyads. Second, privacy-preserving datasets of in-the-wild conversations often omit recording audio. So items referring to the verbal or paralinguistic content of speech were skipped, thereby relying solely on nonverbal cues for perception. Finally, we excluded original items that would require external raters to make significant speculations about participants' desires and opinions beyond what can be inferred from their observable behavior. These include questions related to interpersonal liking (e.g. \textit{``I would like to interact more with the partner in the future''}), or degree of rapport (e.g. \textit{``I felt that the partner was paying attention to my mood''}). From the varied descriptions of pilot judges on the matter, as well as internal author discussions, we deemed that answering such questions require external observers to make too many unverifiable assumptions for a useful perceived measure of conversation quality. We provide the two PCQ questionnaires in Supplementary Material Section 1.

\section{Annotations, Validity, and Reliability}\label{Section:Annotations}

\subsection{Dataset}\label{Sec:Dataset}
We use the publicly available MnM dataset \cite{cabrera2018matchnmingle}. MnM is a multimodal dataset of in-the-wild free-standing mingling interactions. The recordings constitute a total of $30$ minutes of interaction across three days, annotated for conversation groups using the spatial positions of the participants in video from overhead cameras. \figurename~\ref{fig:mnm} illustrates a snapshot from the dataset. Conversation groups were operationalized using the framework of F-formations \cite{kendon1990conducting}. The authors chose specific windows of $10$ minutes per day for annotation with an aim to eliminate possible effects of participant acclimatization to being in a recorded mingling setting, and to maximize the density of participants in the scene. Over the $30$ minutes $174$ conversation groups were annotated. The duration of group conversation follows a mean of $1.91$~min, std. of $2.13$~min, median of $1.10$~min, and a mode of $0.52$~min. The provided data contains video from three of the five overhead cameras, and accelerometer readings from a sensor pack worn by each participant.

\subsection{Annotation Procedure}

\begin{figure}[t]
  \begin{subfigure}[]{0.49\columnwidth}
    \centering
    \includegraphics[width=\linewidth]{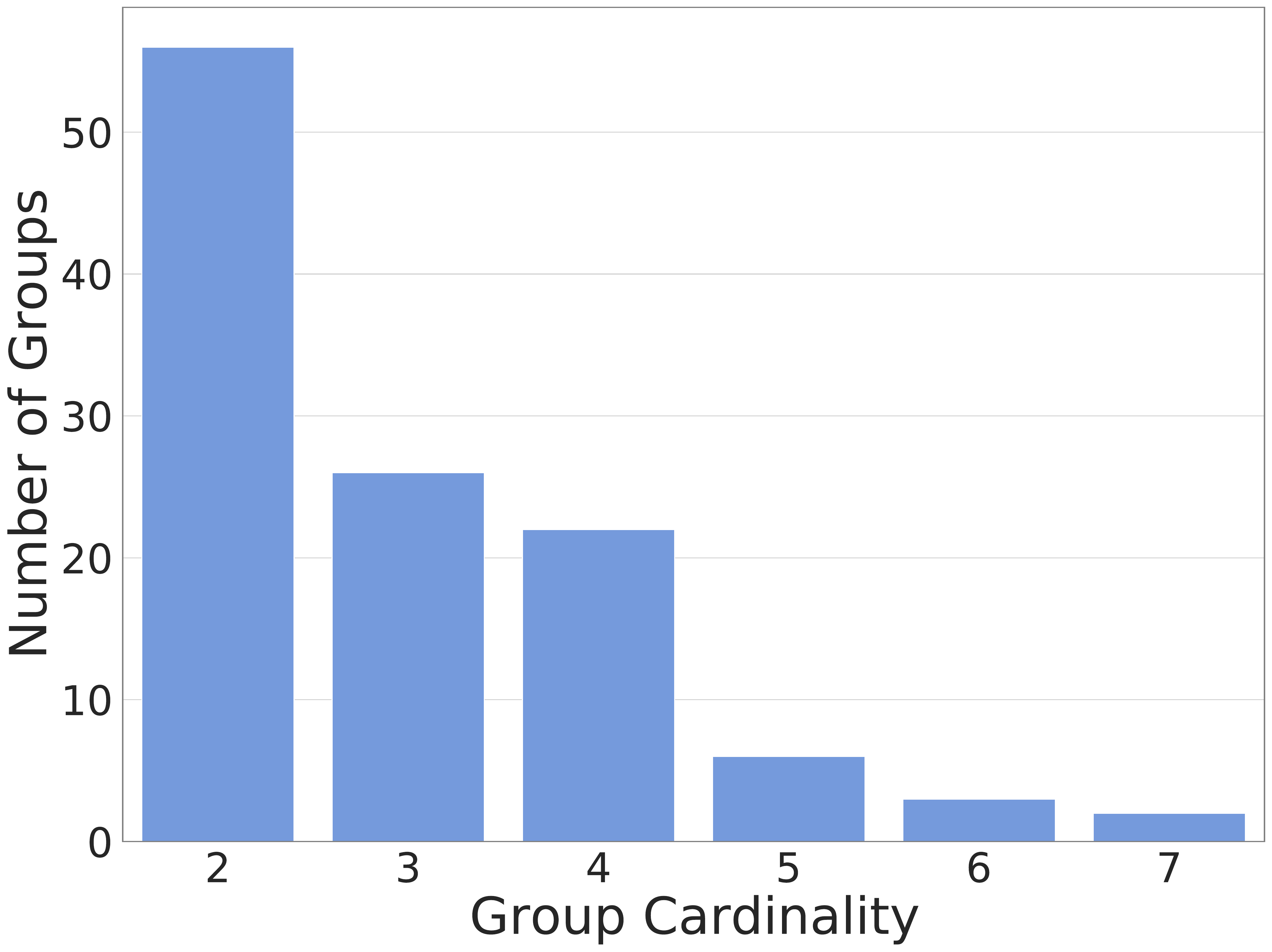}
    \caption{Group Cardinality}
    \label{fig:fform-dist-size-final}
  \end{subfigure}
  \hfill %%
  \begin{subfigure}[]{0.49\columnwidth}
    \centering
    \includegraphics[width=\linewidth]{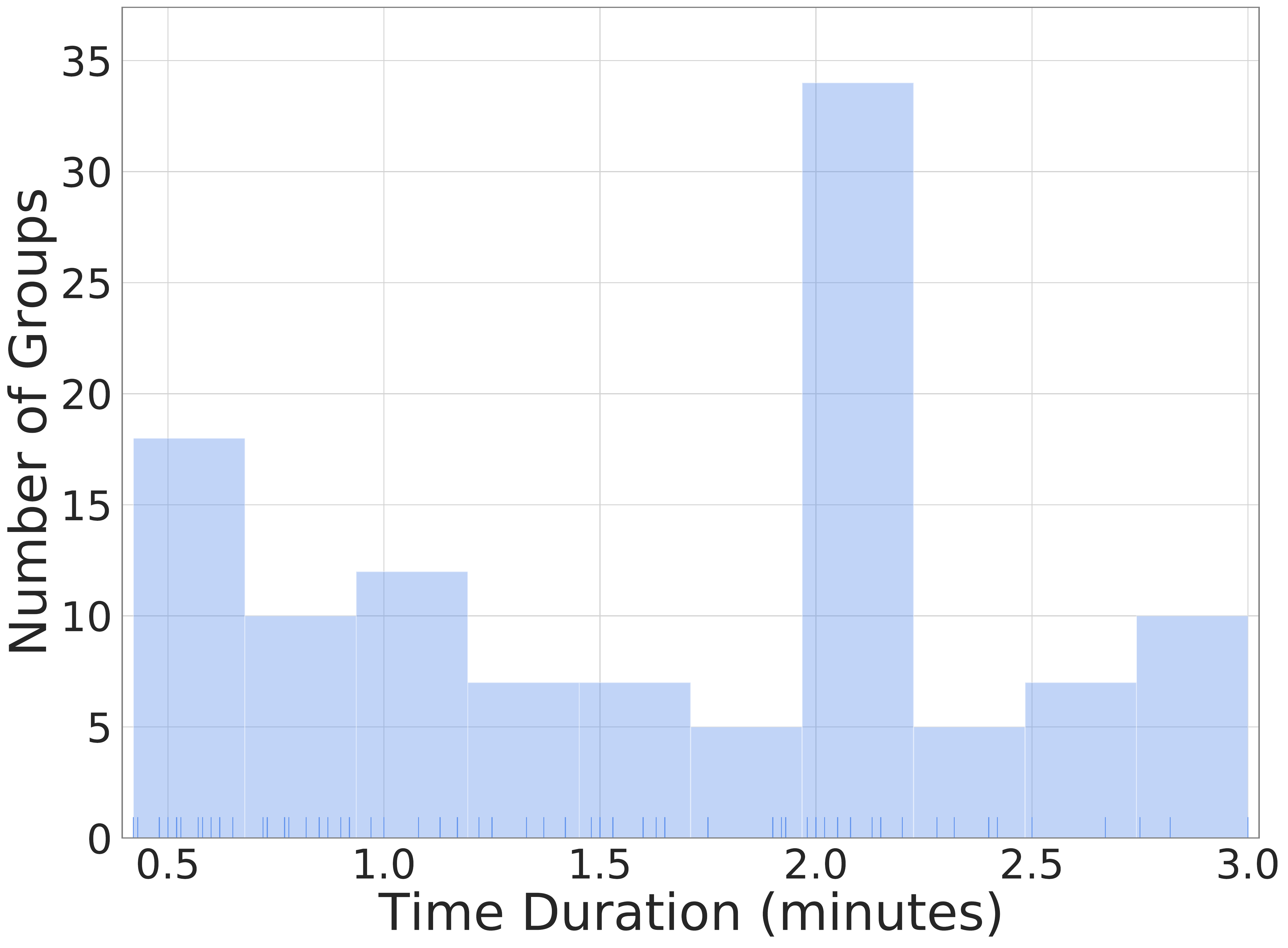}
    \caption{Duration of interactions}
    \label{fig:fform-dist-duration-final}
  \end{subfigure}
  \caption{Distribution of conversation group attributes from the MatchNMingle dataset.}
\end{figure}

Our annotations for PCQ were performed using only the video from the overhead cameras. The available data only contains ambient audio from the overhead cameras that isn't sufficient to infer verbal content reliably, and close-talk microphones were avoided to preserve privacy. However, video recordings capture rich non-verbal behaviours of participants from which a useful perception of conversation quality can be formed \cite{Lindley2013, gatica_interest}.

We began by splitting the group conversations into multiple thin-slices \cite{Jaques2016, murphy2015reliability}. Considering the distribution of group interaction duration in the data, we split conversations of duration greater than $2$ minutes into independent slices of $1$ minute each. Conversations of duration less than $2$ minutes were kept as is. We also omitted groups with duration less than $30$~seconds. After the omission, the total number of resulting conversation groups was $115$. The distribution of group cardinality (number of participants) and interaction duration can be seen in \figurename~\ref{fig:fform-dist-size-final} and \figurename~\ref{fig:fform-dist-duration-final} respectively. 

We began by first conducting a qualitative annotation pilot with the same na\"ive judges who participated in the qualitative interviews. Note that these judges were not used for the final annotations. The goal of this pilot was to fine-tune the final annotation process using any initial feedback about the annotation procedure. The pilot annotators were presented with the videos of the individual thin-slices and asked to fill the two PCQ questionnaires. However, post-hoc interviews revealed two considerations. First, the annotators found the presence of free-standing conversation groups (FCGs) other than the one under consideration distracting. Second, the annotators suffered from fatigue while annotating longer conversations, especially while annotating both individual and group level PCQ. In light of this feedback, we cropped each FCG from the overhead video. To further reduce fatigue, annotators were given a period of two months to annotate all the slices, and were instructed to not annotate more than three groups per day.

% We began by cropping each annotated conversation group from the overhead video. This was done in order to prevent annotators from getting distracted away from the specific group in focus. Next, interactions were split into multiple thin slices of interactions and then presented to annotators as independent clips of social interactions. This was done in order to collect more reliable and granular annotations for longer group interactions. 

% From Figure-\ref{fig:fform-dist-duration}, we see that the durations of f-formation interactions varies widely, from interactions of few seconds to that greater than 3-4 minutes. In that case, it is not reliable enough to have only one label annotation to define the conversation quality for the f-formation interactions of different durations.

% In this subsection, we explain the annotation procedure used to collect annotations for perceived \textit{Conversation Quality}. While explaining the annotation procedure, we also discuss several key considerations taken to devise the strategy.

% The video clips of the spontaneous interactions, filmed using overhead cameras, was the only modality used for the manual annotation of the \textit{Conversation Quality}. No audio data was used for the annotation process.

% Moreover, annotations using audio recordings are also time consuming as they are generally prone to language constraints, noise and talk-overlap.  annotations using only video recording are resource efficient.

The final annotations\footnote{Annotations will be available on the MathcNMingle website at \url{http://matchmakers.ewi.tudelft.nl/matchnmingle/pmwiki/}} were performed on a five-point scale by three annotators, aged between $22$ and $30$ years, two female and one male. The annotators were also chosen to be na\"ive judges in order to capture a general perception of conversation quality. The annotators were provided with the independent conversation slices of cropped video clips and asked to fill out both PCQ questionnaires. The slices were provided to the annotators in randomized order for each annotator, to prevent any annotator bias which might occur from a chronological ordering of the clips.

\subsection{Validity}

When measuring intangible constructs such as PCQ, it is important to assess the validity \cite{questionnaire_brinkmann, cronbach1955construct} of the proposed instrument. Broadly, validity deals with whether the instrument indeed measures what it claims to be measuring. 

\subsubsection{Face Validity}
First we tested the face validity of our questionnaire items. Face validity is a consensus measure, and is checked to ensure that the raters accept the instrument \cite{questionnaire_brinkmann}. This is done by asking the raters if the items seem valid. Both questionnaires passed the face validity test with full consensus. 
    
\subsubsection{Criterion and Construct Validity}
When prior trusted sources or standards exist for a construct, a criterion-oriented study is common. Here validity can be established by showing that results of administering the instrument correlates with a contemporary criterion (e.g. a psychiatric diagnosis) or by proposing one instrument as a substitute for another (e.g. a multiple-choice form of spelling test is substituted for taking dictation) \cite{cronbach1955construct}. However, since PCQ is a novel conceptualization, prior trusted sources and standards don’t exist for it. In such cases where the attribute being measured is not ``operationally defined'', construct validity must be investigated \cite{cronbach1955construct, questionnaire_brinkmann}. Construct validation is the gathering of evidence to support the interpretation of what a measure reflects, and addresses the question ``What constructs account for variance in test performance?''

\begin{figure}[t]
  \begin{subfigure}[]{0.49\columnwidth}
    \centering
    \includegraphics[width=\linewidth]{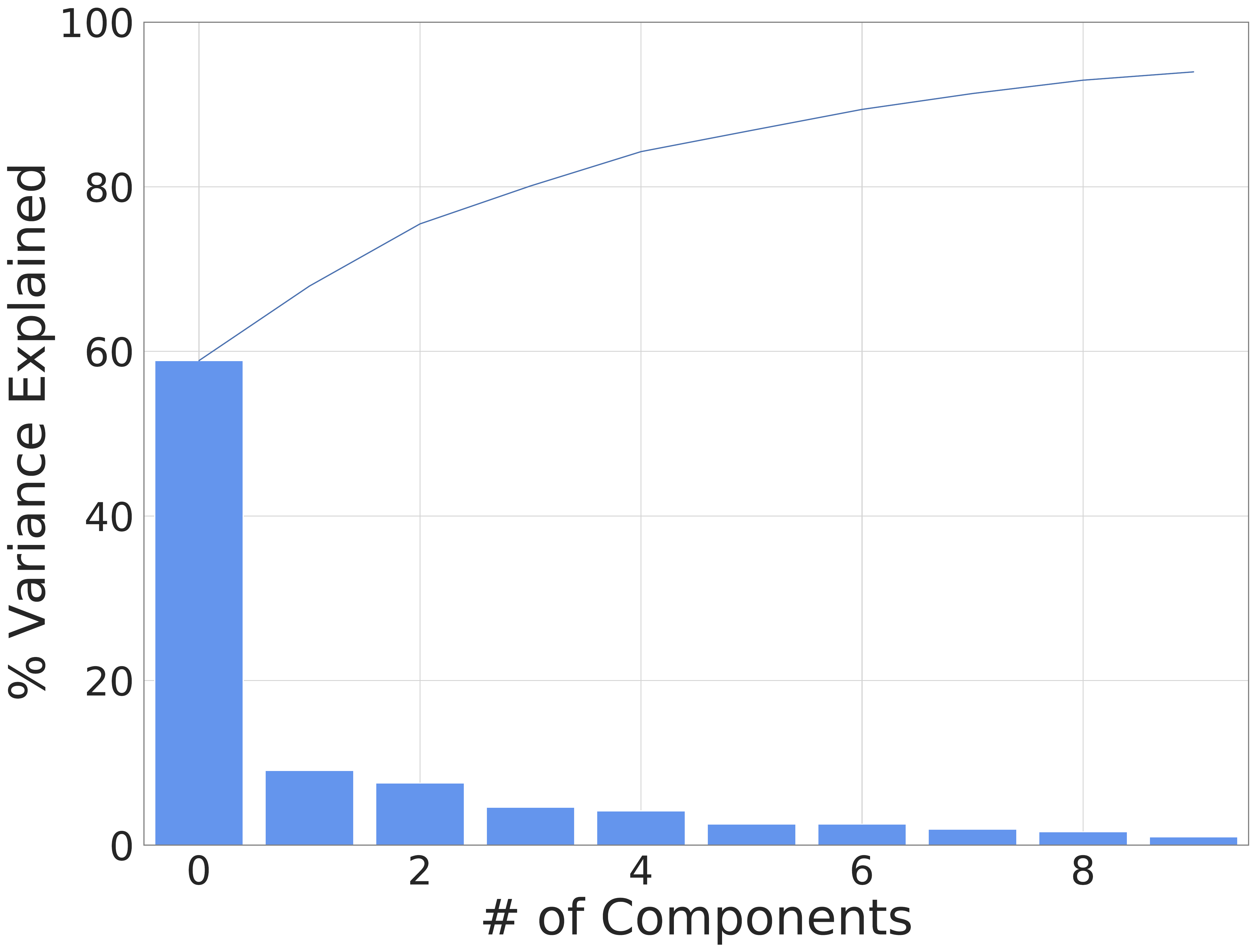}
    \caption{Group-level}
    \label{fig:eigen-group}
  \end{subfigure}
  \hfill %%
  \begin{subfigure}[]{0.49\columnwidth}
    \centering
    \includegraphics[width=\linewidth]{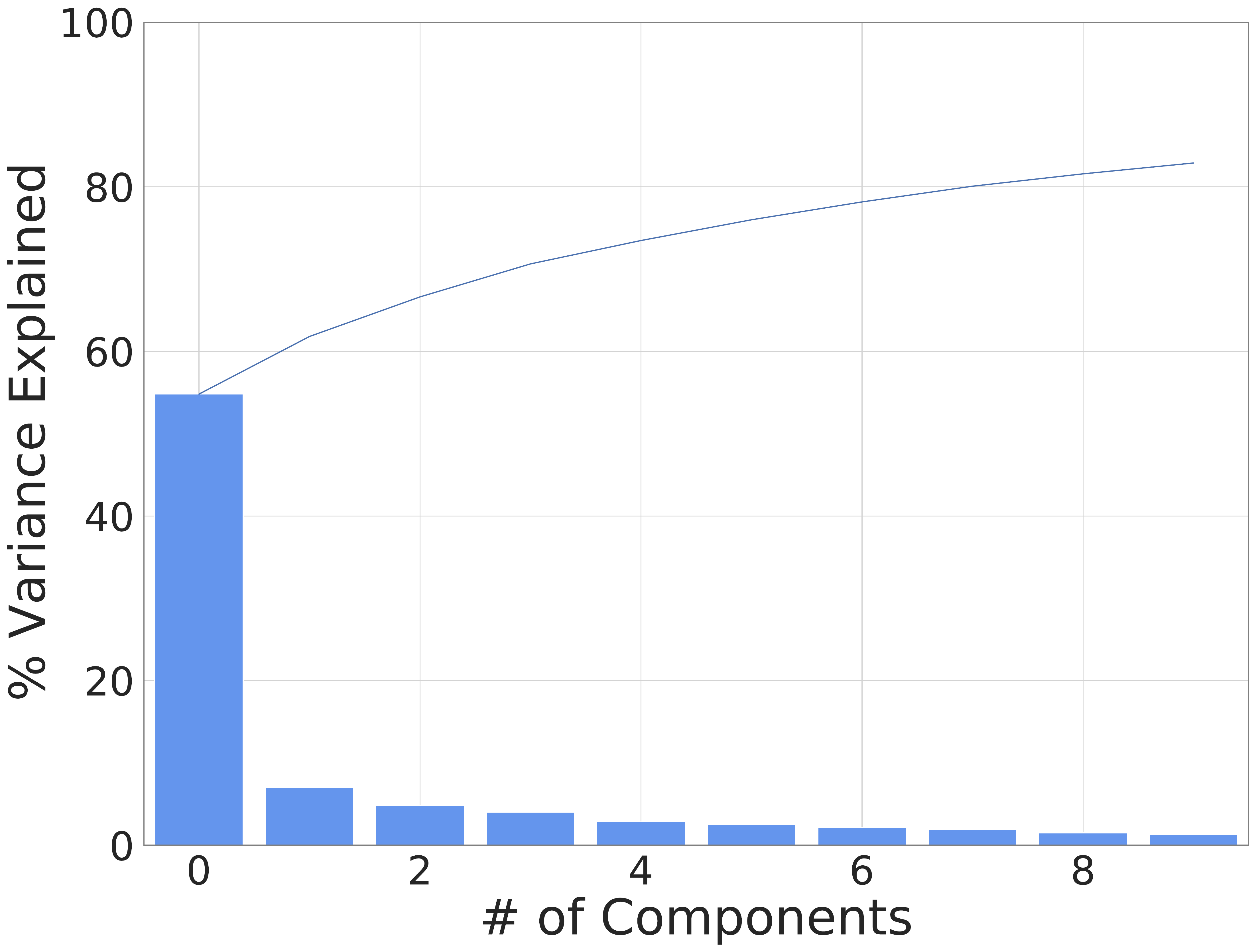}
    \caption{Individual-level}
    \label{fig:eigen-indiv}
  \end{subfigure}
\caption{Eigenvalue distribution (bar chart) and the cumulative percentage of the explained variability (line plot).}
\label{fig:eigen-dist}
\end{figure}
    
\begin{figure}[b]
  \begin{subfigure}[]{0.49\columnwidth}
    \centering
    \includegraphics[width=\linewidth]{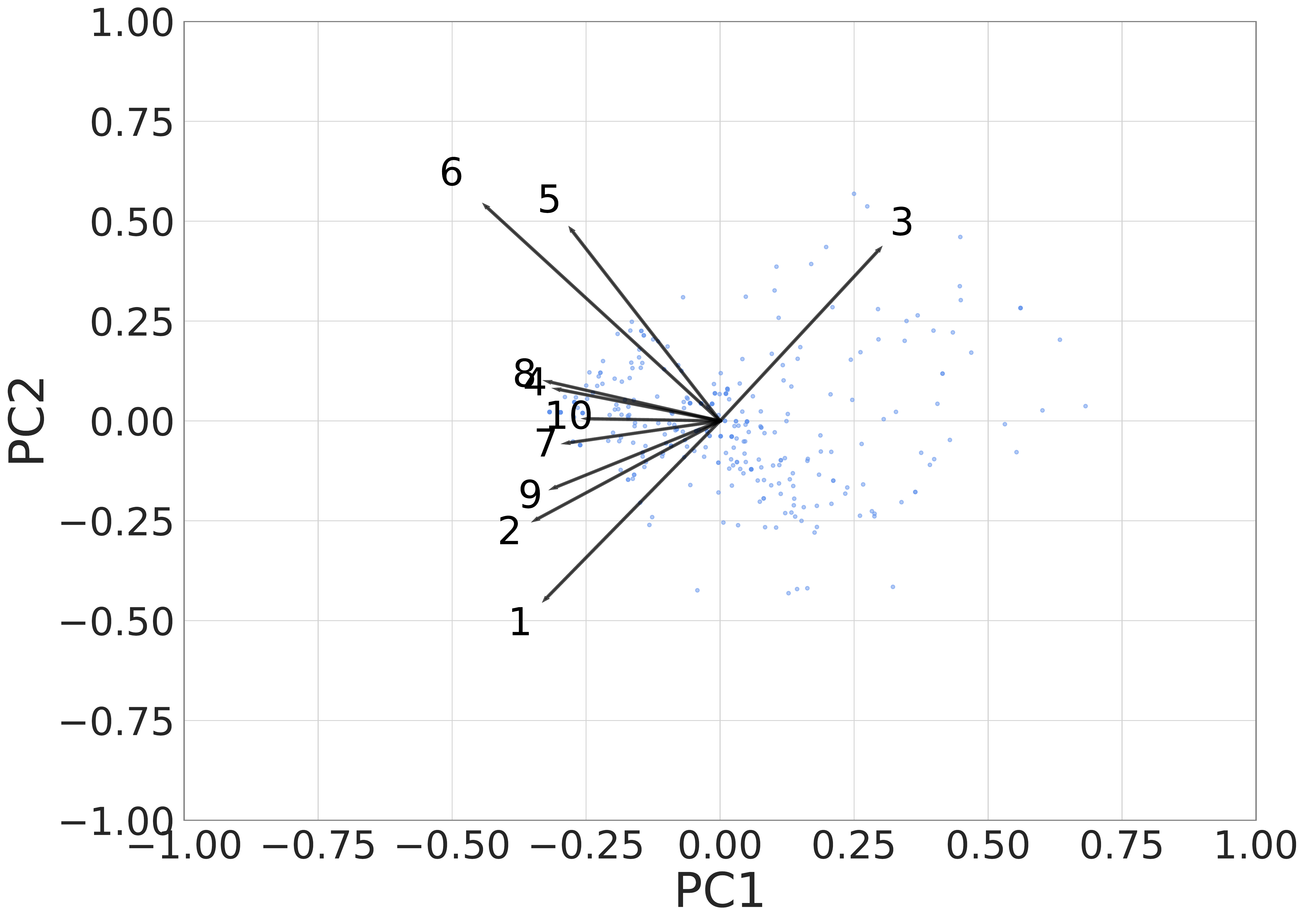}
    \caption{Group-level}
    \label{fig:load-group}
  \end{subfigure}
  \hfill %%
  \begin{subfigure}[]{0.49\columnwidth}
    \centering
    \includegraphics[width=\linewidth]{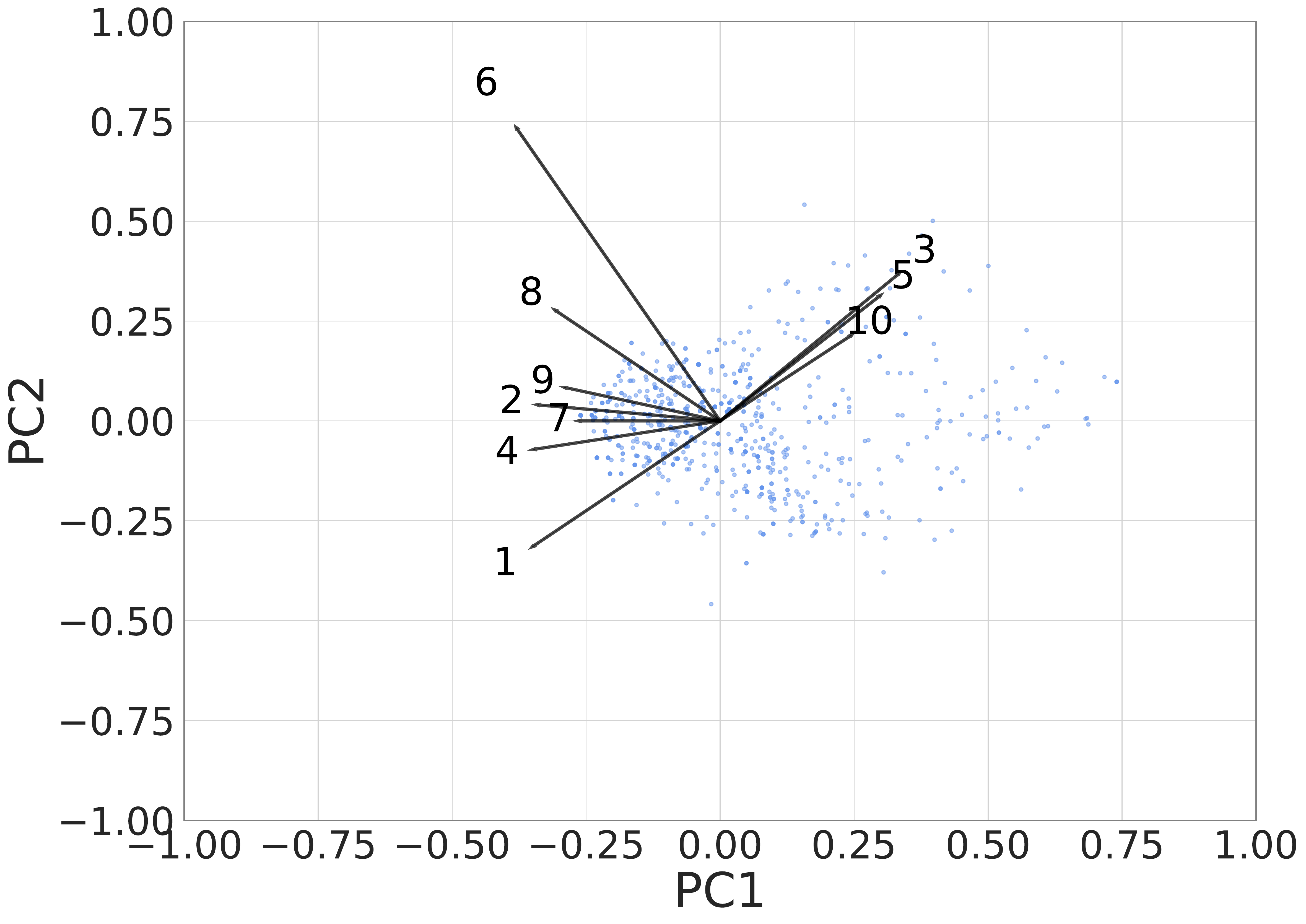}
    \caption{Individual-level}
    \label{fig:load-indiv}
  \end{subfigure}
\caption{Plot of the factor loadings (black lines) and the samples (blue dots) in the first two principal components.}
\label{fig:loading-dist}
\end{figure}

A typical approach for construct validation involves performing a factor analysis and investigating if items corresponding to one construct correlate with each other along a factor (convergent validity) and divert from items of other constructs (divergent validity) \cite{questionnaire_brinkmann}. This works well for instruments with independent constructs (e.g. \textit{gender} and \textit{complexity of use} in \citeauthor{questionnaire_brinkmann}'s mobile phone design questionnaire \cite[Table~$9$]{questionnaire_brinkmann}). However, such an analysis is unsuitable for situations like ours with overlapping constructs. Indeed, \citet{Cuperman2009} decided to not reduce items from the POI to a smaller set of factors, following a precedent set by \cite{funder1993behavioral}. In contrast, we do perform a factor analysis, but rather than seeking the independence of factors, we investigate whether the loadings correspond to interpretable attributes of the constructs. 

A principal component analysis of the annotations showed that $71\%$ and $65.2\%$ of the variance at the group-level and individual-level respectively could be explained by the first principal component (see \figurename~\ref{fig:eigen-dist}). From the plot of the data samples using the first two principal components in \figurename~\ref{fig:loading-dist}, we see that questions corresponding to positive and negative orientations of PCQ cluster in opposite directions along the two components. Specifically the individual-level items pertaining to awkwardness ($3$), discomfort ($5$), and self-consciousness ($10$) load in the exactly opposite direction to the item about the individual looking relaxed ($1$). Of these, at the group-level only items $1$ and $3$ apply, and we see a similar pattern. Further, we also observe that the items pertaining to \textit{equal opportunity} cluster separately: these correspond to items $5$ and $6$ about free-for-all participation at the group-level, and item $6$ about taking lead at individual-level. Specifically, the highest loading of individual-level item $6$ suggests that the attribute of taking lead in conversations it accounts for the highest variance between individuals, which is intuitive given prior work on dominance in groups \cite{jayagopi2009modeling}.

% \subsubsection{Criterion Validity} With criterion validity, results of a questionnaire are compared against other trusted sources or standards known to measure the construct \cite{questionnaire_brinkmann}. However, since PCQ is a novel conceptualization, prior trusted sources and standards don't exist for it. Nevertheless, we adapt our proposed questionnaire items from existing instruments that we believe to be valid for the constructs they measure, following an approach similar to other works \cite{Jaques2016, Cerekovic2014}.

\subsection{Reliability}

\begin{figure}[t]
\begin{subfigure}[]{0.49\columnwidth}
    \centering
    % \captionsetup{justification=centering}
    \includegraphics[width=\linewidth]{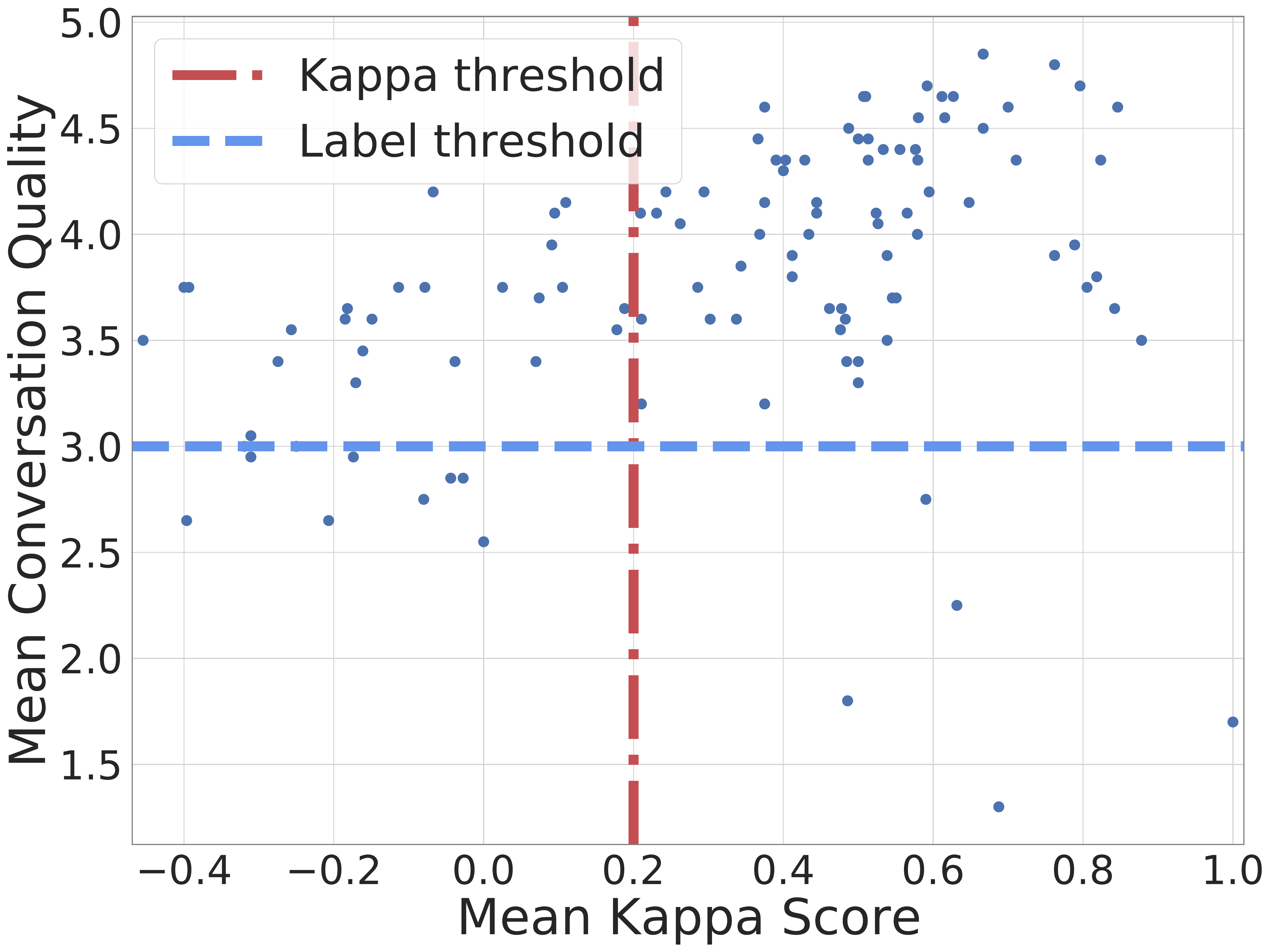}
    \caption{Group-level}
    \label{fig:all3-kappvsconvq-group}
  \end{subfigure}
  \hfill %%
  \begin{subfigure}[]{0.49\columnwidth}
    \centering
    % \captionsetup{justification=centering}
    \includegraphics[width=\linewidth]{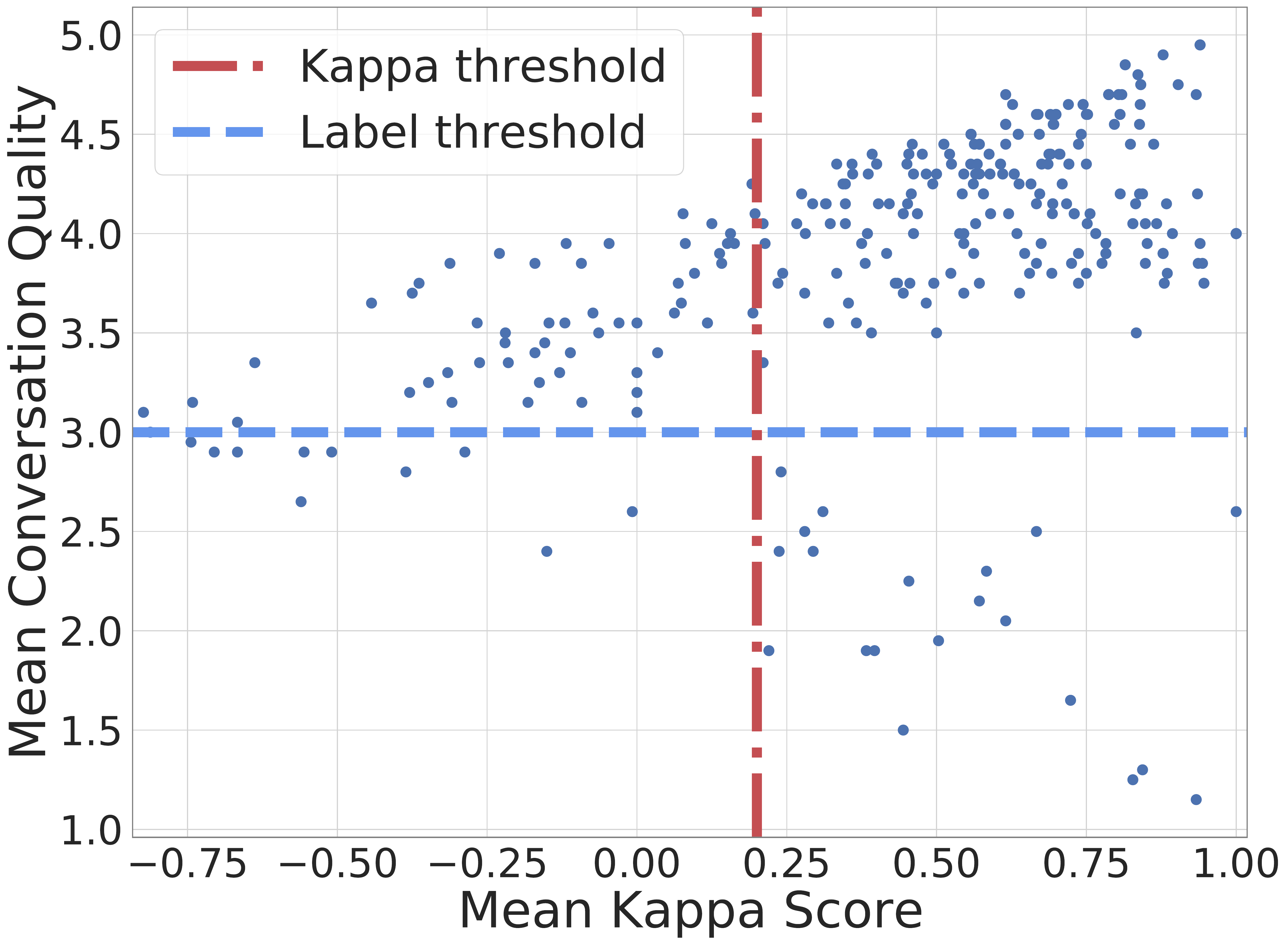}
    \caption{Individual-level}
    \label{fig:all3-kappvsconvq-indiv}
  \end{subfigure}
\caption{Scatter plot of the Mean Kappa score ($\kappa$) vs the Mean Conversation Quality score.}
\label{fig:all3-kappvsconvq}
\end{figure}

For estimating inter-annotator agreement, we use the quadratic weighted kappa measure ($\kappa$) \cite{cohen1968weighted}, a variant of the Cohen's kappa. The measure is especially useful when the annotation data is ordinal in nature. In \figurename~\ref{fig:all3-kappvsconvq} we plot the mean kappa score against the mean conversation quality score in a scatter plot similar to the analysis of inter-annotator agreement for cohesion performed by \citet{Behavior2010}.

From the plots we see that there exists a linear relationship between mean kappa scores and mean conversation quality scores, suggesting that annotators agree better on conversations of higher quality than conversations of lower quality. Moreover, in the individual-level annotations, there exists a small cluster of samples where annotators tended to agree higher for lower conversation quality samples as well. In contrast, annotators never agree well for low conversation quality samples at the group-level. 

% But, this was not expected by us. We expected similar results as seen in Hung et al.'s work \cite{Behavior2010}, where inter-annotator agreements on cohesion levels for meetings were higher at the two extremes of the scale. Such a behaviour is seen only marginally and only for the individual-level annotations.

    % \begin{figure}[ht]
    % \begin{subfigure}[ht]{0.5\columnwidth}
    %     \centering
    %     \includegraphics[width=\linewidth]{images/annotation/All3-Group-ZM.png}
    %     \caption{Group-level annotations.}
    %     \label{fig:all3-kappvsconvq-group-zm}
    %   \end{subfigure}
    %   \hfill %%
    %   \begin{subfigure}[ht]{0.5\columnwidth}
    %     \centering
    %     \includegraphics[width=\linewidth]{images/annotation/All3-Indiv-ZM.png}
    %     \caption{Individual-level annotations.}
    %     \label{fig:all3-kappvsconvq-indiv-zm}
    %   \end{subfigure}
    % \caption{Scatter plot between Mean Kappa score ($\kappa$) and Mean Conversation Quality score, after ZM adjustment.}
    % \label{fig:all3-kappvsconvq-zm}
    % \end{figure}

% Chirag Raman: Say RECOLA uses 0.2 and above, Hung and perez reject 0.3 and below, and we follow Hung and Perez

To handle low inter-annotator agreement, following suggestions by \citet{ringeval2013introducing}, we performed zero-mean local normalization to remove annotator bias. \citet{Behavior2010} omit samples below $\kappa=0.3$, and \citet{ringeval2013introducing} obtain an average $\kappa$ of $\approx0.2$ for all their emotion dimensions. Following these approaches, data samples at both the group- and individual- levels with $\kappa<0.2$ were omitted from further analysis, where a $\kappa>0.2$ indicates a reliability of \textit{fair and above} \cite{mchugh2012interrater}.
 
%  We performed similar ZM adjustments on our annotations and similar analysis was performed. 
 
%  The resulting plots, seen in Figure-\ref{fig:all3-kappvsconvq-zm}, shows that no major changes are seen post the ZM technique. This suggests that there exists no mean shift between annotators but there exists a basic difference in annotator judgements. 

\section{Modeling Conversation Quality}
In this section we describe the experimental setup for our study of behavioral features that can be predictive of PCQ. 

\label{Sec:ModelingConvQ}

% \CR{Why is this the first paragraph?}
% The MatchNMingle dataset used tri-axial accelerometers, worn by all participants, to record their bodily movements. The accelerometer recorder bodily movements along 3-direction (3 dimensions), specifically, along the X axis - capturing the horizontal (left-right) movements, along the Y axis - capturing the vertical (up-down) movements and along the Z axis - capturing the 3rd dimensional forward-backward movements. The tri-axial accelerometer recorded the bodily movements at a sampling frequency of 20Hz, recording 20 fine-grained samples every second. 

% In this section, we present the techniques involved in the statistical analysis and the predictive modeling of individual- and group- level conversation quality measures. 
% Firstly, we present the technique used to preprocess the raw tri-axial accelerometer signals. Subsequently, we present the features extracted from the preprocessed signals. Finally, we explain in detail the experimental setup and strategy deployed for the study of conversation quality.

\subsection{Preprocessing}
We first preprocess the raw tri-axial acceleration signal from the wearable sensors to extract low-level features. First, each axis recording from the tri-axial accelerometer is standardized by calculating the z-score for each individual and axis, thereby removing the individual differences in movement intensity. Following prior work using wearable sensor data for the study of conversation dynamics \cite{hung2013classifying, gedik2016speaking, kapcak2019estimating}, we compute the following features using the z-scores: the raw and absolute values for $3$ axes each, and the Euclidean norm of the raw values across axes, resulting in a total of $7$ feature channels. Further, we denoise the feature channels following the common practice of extracting channel-wise statistical and spectral features using sliding-windows, which smooths the signal. In the following experiments, we include both the noisy and the denoised features to circumvent any data loss from the denoising. 

\subsection{Feature Extraction}

\begin{table}[t]
    \caption{An overview of the four sets of individual- and pair- level behavioural features extracted.}
    \centering
    \begin{tabular}{lll} 
    \toprule
    & \textbf{Attribute Category} & \textbf{Attribute Variant} \\
    \midrule 
    \multicolumn{3}{l}{\textbf{Synchrony}}\vspace{4pt}\\
    1 & Correlation             & correlation coefficient ($\rho_{xy}$) \\
    2 & Time-lagged Correlation & min, max, argmin, argmax \\
    3 & Mutual Information      & min, max, mean, variance \\
    4 & Mimicry                 & lag\_min, lag\_max, \\ 
                                & & lag\_mean, lag\_variance, \\ 
                                & & lead\_min, lead\_max, \\
                                & & lead\_mean, lead\_variance\\
    \midrule
    \multicolumn{3}{l}{\textbf{Causality}}\vspace{4pt}\\
    5 & Coherence               & min, max \\
    6 & Granger's Causality     & f\_value \\ 
    \midrule
    \multicolumn{3}{l}{\textbf{Convergence}}\vspace{4pt}\\
    7 & Symmetric Convergence   & $\rho$ \\
    8 & Asymmetric Convergence  & lag, lead \\
    9 & Global Convergence      & $d_1 - d_2$ \\
    \midrule
    \multicolumn{3}{l}{\textbf{Turn-Taking}}\vspace{4pt}\\
    10 & Conversation Equality  & degree of equality \\
    11 & Conversation Fluency   & percentage of silence,\\
                                & & \# back-channels \\
    12 & Conversation Synchronisation   & percentage of overlap, \\
                                        & & \# successful interrupts, \\
                                        & & \# unsuccessful interrupts \\
    \bottomrule
    \end{tabular}
    \label{tab:feats-overview}
\end{table}

We consider bodily coordination features and turn-taking based features to study PCQ. For bodily coordination, we extract three sets of features: \textit{synchrony, convergence,} and \textit{causality}. An overview of the individual and pairwise features extracted can be seen in Table~\ref{tab:feats-overview}.

\textbf{Synchrony:} Synchrony estimates the dynamic and reciprocal adaptation of the temporal structure of behaviors between interlocutors \cite{Delaherche2012}. Following existing literature \cite{Nanninga2017, kapcak2019estimating, Hagad_rapport}, we extract four unique measures of interpersonal synchrony: \textit{Correlation}, \textit{Time Lagged Correlation}, \textit{Mutual Information}, and \textit{Mimicry}.

\textbf{Causality:} Correlation does not adequately capture the causal effect \cite{aldrich1995correlations}. We therefore extract two causality features: \textit{Coherence} \cite{richardson2005looking} and \textit{Granger's Causality}.

\textbf{Convergence:} These features capture the increasing similarity between interacting partners over time \cite{edlund2009pause}, and have been shown to be predictive of mutual liking, attraction \cite{kapcak2019estimating, michalsky2017pitch}, and social cohesion \cite{Nanninga2017}. In this research, we use three unique estimates of convergence: \textit{Symmetric Convergence}, \textit{Asymmetric Convergence}, and \textit{Global Convergence}.

\textbf{Turn-Taking:} MnM provides binary speaking status of participants annotated from video data. We extract turn-taking features using these annotations by assuming a speaking turn to be a continuous speaking activity segment separated by at least $500$~ms of silence \cite{lai2013modelling, Lindley2013}. Following existing literature \cite{lai2013modelling, Lindley2013, Behavior2010}, we extracted turn-taking features under three categories: \textit{Conversation Equality}, \textit{Conversation Fluency}, and, \textit{Conversation Synchronisation}. Assuming a conversation of duration $T$ and a group of $N$ people, and denoting the $i$-th individual's binary speaking status as $\bm{s}^i = [s^i_1, \ldots, s^i_T]$, we have the precentage of speaking duration for $i$, $d_{\mathrm{speak}}^i = (\sum_{t \in [T]} \bm{s}^i_t) / T$.   
We compute degree of equality for $i$ as $eq^i = (d_{\mathrm{speak}}^i - \bm\bar{d} )/\bm\bar{d}$, where $\bm\bar{d} = (\sum_{i\in[N]} d_{\mathrm{speak}}^i)/N$. As measures of fluency, we compute the percentage of individual silence $d_{\mathrm{silence}}^i = 1 - d_{\mathrm{speak}}^i$ and the number of back-channels (very short utterances of duration up to $2$~s). As a measure of synchronisation, we consider the percentage of speech overlap, computed for the $i$-th individual as $d_{\mathrm{o}}^i = (\sum_{t \in [T]} \mathbbm{1}\{s_t^i = s_t^{j, j \neq i}\})/T$, and the number of successful and unsuccessful interruptions \cite{Behavior2010}. 

Finally, following \cite{Nanninga2017, Zhang2018_TeamSense}, we translate individual and pairwise features to group-level features using the feature aggregates \textit{minimum}, \textit{maximum}, \textit{mean}, \textit{mode}, \textit{median} and \textit{variance}. Specifically, for individual-level modeling, similar to \citet{rapport_muller} we aggregate over pairwise features involving that particular individual, and for group-level modeling aggregation is done over all the pairs in the group.

\subsection{Experimental Setup}

\subsubsection{Statistical Analysis}

We perform hypothesis-driven tests to study the effect of (i) group cardinality, (ii) turn-taking attributes and (iii) body coordination attributes on PCQ. We use the Quantile Least Squares (QLS) and Joint LASSO models for our hypothesis-driven analysis. The QLS analysis considers each set of behavioral features independently, while the Joint LASSO analysis accounts for the combined effect of all features, allowing for complementary insight.

\textbf{Quantile Least Squares:} 
QLS fits the regression to the conditional \emph{median} of the dependent variable, in contrast to the conditional \emph{mean} estimated by Ordinary Least Squares (OLS). Intuitively, the conditional median is more robust against outliers. More crucially, the QLS does not require the data to abide the assumptions of exogeneity and homoscedasticity like the OLS does. We find that the variance of the independent variables varies largely across quantiles (see sSupplementary Figure~S1 for scatter-plots), thereby violating the exogeneity and homoscedasticity assumptions. We therefore use the QLS model for our analysis.
    
\textbf{Joint LASSO:} While QLS is convenient in situations where classical parametric assumptions do not hold, it still suffers from effects of multicollinearity. We therefore use the QLS model to study behavioural feature sets in isolation. However, to also account for the combined effect of feature sets, we perform a \emph{joint} regression over all features using a LASSO model which induces sparsity to address multicollinearity. For further insight, we perform a post-hoc Spearman's rank correlation on the LASSO filtered features. 

An overview of the of statistical tests performed can be seen in Table \ref{tab:stat-test-structure}. We denote individual- and group- level PCQ as IndivPCQ and GroupPCQ respectively. In total, with two dependant variables, three sets of independent variables and three statistical models, $18$ tests were performed. Bonferroni correction is applied to the p-values to correct for multiple testing for each dependent variable. After Bonferroni correction a significance threshold of $0.005$ was used for testing significance in all the analyses presented.

\begin{table}[t]
    \caption{Overview of the statistical analysis performed.}
    \centering
    \begin{tabular}{ccc} 
        \toprule
        \textbf{\begin{tabular}[c]{@{}c@{}}Dependent\\ Variables\end{tabular}} & \textbf{\begin{tabular}[c]{@{}c@{}}Independent\\ Variable Sets\end{tabular}} & \textbf{\begin{tabular}[c]{@{}c@{}}Statistical\\ Models\end{tabular}} \\
        \midrule 
        IndivPCQ  & Group cardinality  & QLS Regression  \\
        GroupPCQ  & Turn-talking       & LASSO Regression \\
                 & Bodily Coordination & Rank Correlation \\
        \bottomrule
    \end{tabular}
    
    \label{tab:stat-test-structure}
\end{table}

% The
\subsubsection{Analysis of Feature Extraction and Fusion}

We perform data-driven analyses to study the effects of (i) window sizes for data preprocessing; (ii) fusion of attribute categories; and (iii) feature aggregators to compute group-level features from individual and pairwise features. 

For these analyses, we treat predicting PCQ as a binary classification of low and high PCQ scores. A threshold of $3.0$ (on the $5$-point scale) is used to binarize the scores into low and high. As such, our annotations suffer from class imbalance (see \figurename~\ref{fig:all3-kappvsconvq} for the label threshold). To address this, we employ the Synthetic Minority Oversampling technique (SMOTE) \cite{chawla2002smote}, which generates synthetic samples from the minority class. We use a logistic regression model trained with the elastic loss that combines the $L_1$ and $L_2$ penalties of the lasso and ridge regularization methods. Specifically, for each experiment we evaluate how the feature extraction or aggregation affects the predictive capability of the model. We jointly consider applicable feature sets, and perform z-score standardization and PCA for dimensionality reduction as preprocessing steps. As performance metrics, we use the confusion matrix, the ROC curve, and the AUC score. The metrics are calculated as the average across $5$-folds in the cross-validation setting. Code for all experiments and analyses are available at \url{https://github.com/LRNavin/conversation_quality}.

\section{Results}
\label{Sec:Exp}
\subsection{Statistical Analysis}
\subsubsection{Analysis of Group Cardinality}
Existing research \cite{gedik2018detecting, sacks1978simplest, Raman2019} has shown that behaviour in group interactions varies with size of the group (group cardinality). Is this true for PCQ as well? Concretely, we test the hypothesis:

\begin{displayquote}
\textit{For an FCG, the PCQ changes with group cardinality.}
\end{displayquote}

\begin{figure}[t]
 \begin{subfigure}[]{0.49\columnwidth}
        \centering
        \includegraphics[width=.95\columnwidth]{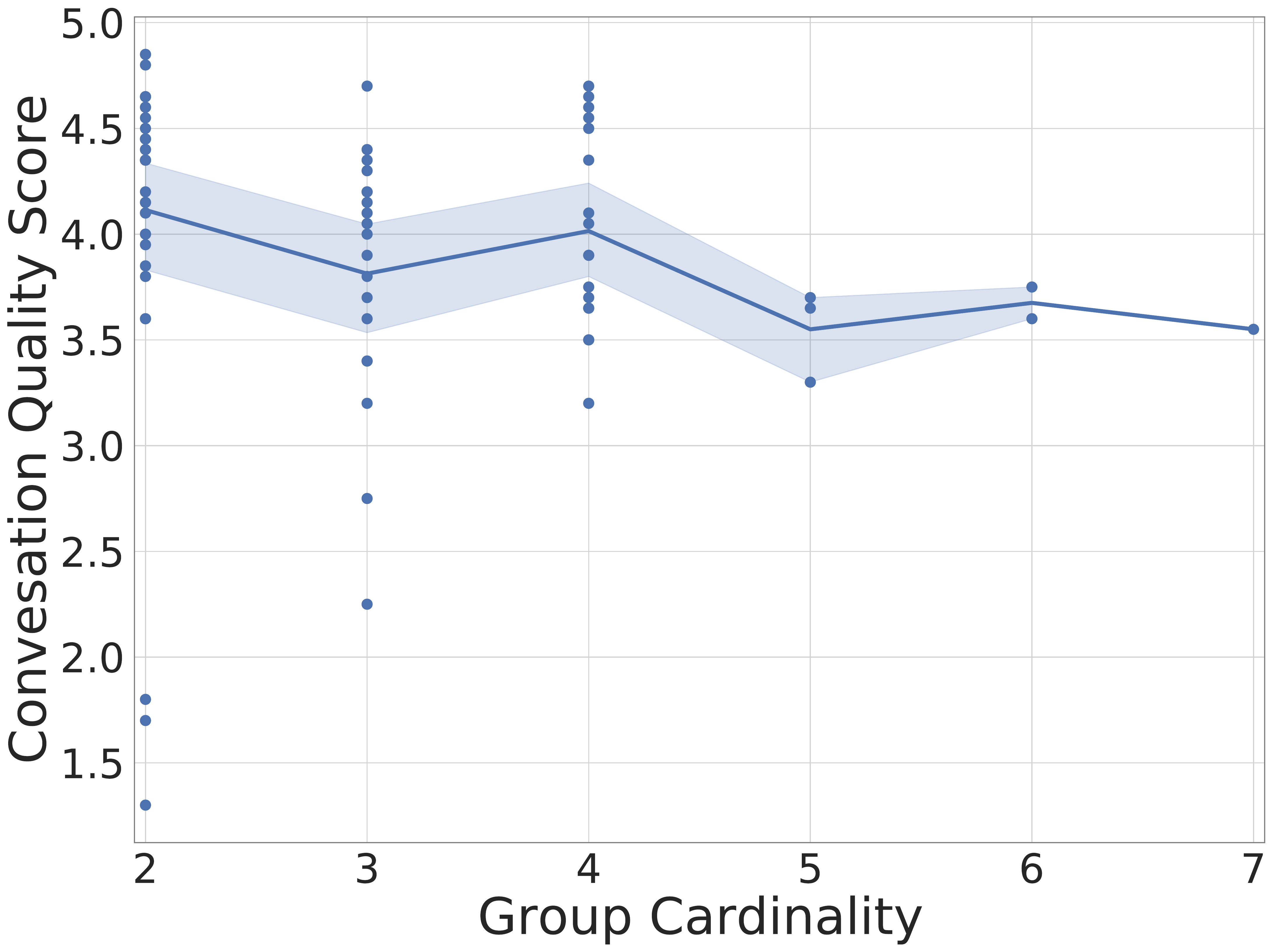}
        \captionsetup{justification=centering, margin=0.5cm}
        \label{fig:indiv-grpsize-swarm}
    \end{subfigure}%
    \begin{subfigure}[]{0.49\columnwidth}    
        \centering
        \includegraphics[width=.95\linewidth]{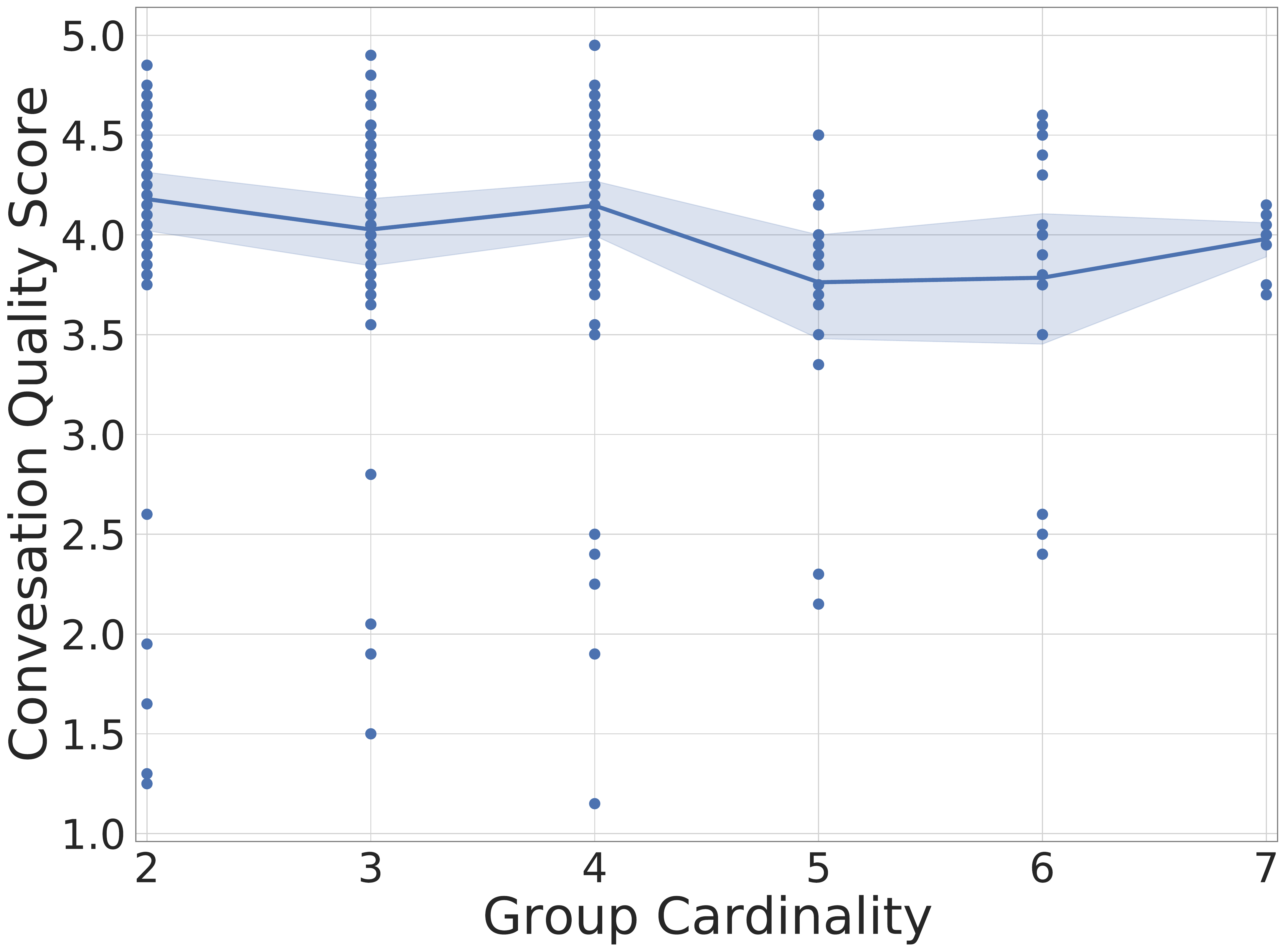}
        \captionsetup{justification=centering, margin=0.5cm}
        \label{fig:indiv-grpsize-scatter}
    \end{subfigure}
    \caption{GroupPCQ and IndivPCQ across group cardinalities.}
    \label{fig:qualitative-indiv-grpsize}
\end{figure}

\noindent From the plots in \figurename~\ref{fig:qualitative-indiv-grpsize}, we see that for both GroupPCQ and IndivPCQ the means for cardinalities of $2$, $3$ and $4$ are higher than that of $5$, $6$, $7$. The statistical tests reveal that that IndivPCQ and GroupPCQ are significantly different across groups of different cardinality. We note that for all regression models, the $\beta$ coefficient for the group cardinality variable is negative, suggesting that PCQ is inversely proportional to group cardinality. For example, the QLS model associates the cardinality attribute with $\beta=-0.2167$ and $\beta=-0.0833$ for IndivPCQ and GroupPCQ respectively (p-value=$10^{-5}$), indicating that people appear to have better quality conversations with fewer partners.

Post-hoc analysis testing for the differences in PCQ between cardinality pairs reveals that the IndivPCQ scores are significantly different in dyadic group interactions when compared to that of interactions in larger groups (cardinality $\geq 3$). One possible alternate explanation of this result is that raters score PCQ more conservatively when there are more partners to pay attention to. Nevertheless, even if this were the case, it would be a valid characteristic of how people perceive behaviors in larger groups. Significant results were not observed for the post-hoc GroupPCQ comparisons, suggesting that no conclusions can be drawn with respect to GroupPCQ regarding pairwise differences with cardinalities. Note that this result should also be interpreted accounting for the small sample size for cardinalities $\geq 5$.

\subsubsection{Analysis of Turn-Taking Attributes}
Turn-taking features have been have shown to be indicative of constructs such as enjoyment and cohesion \cite{Behavior2010, lindley2008social, Lindley2013}. We test the hypotheses:
\begin{displayquote}
\textit{In an FCG, turn-taking attributes (conversation equality, conversation fluency and conversation synchronisation) are positively correlated with PCQ.}
\end{displayquote}

\noindent
For IndivPCQ, the QLS model reveals that conversation equality and percentage of silence are the most significant attributes, with positive ($\beta=0.2136, p=10^{-4}$) and negative ($\beta=-0.5094, p=10^{-4}$) correlations respectively. For GroupPCQ, QLS reveals that the number of successful and unsuccessful interruptions are the the most significant attributes, with negative ($\beta=-0.0859, p=0.001$) and positive ($\beta=0.0956, p=0.002$) correlations respectively. On the other hand, the LASSO and rank correlation models reveal a different set of significant attributes. For IndivPCQ, along with conversation equality and percentage of silence, the two interruption based attributes were also revealed to be significant. Similarly, for GroupPCQ, unlike the QLS, the two interruption attributes are found to be insignificant, while conversation equality, percentage of silence and number of backchannel attributes are found to be significant.

Intuitively, the result implies that observers consider group conversations with more equitable speaking turns and fewer interruptions to be of higher quality. An important thing to note here is that the complementary models associate all attributes with similar trends even though they differ on which attributes they consider to be of statistical significance. Even though the statistical significance of successful and unsuccessful interruptions differ when considered in isolation or jointly with other features, they are associated with negative and positive $\beta$'s respectively, by all models tested. 

\subsubsection{Analysis of Bodily Coordination Attributes}
Coordination features across modalities such as bodily movements \cite{kapcak2019estimating} and paralinguistic speech features \cite{Nanninga2017} have been shown to be indicative of liking, attraction, and cohesion. Here we test the hypothesis:
\begin{displayquote}
\textit{In an FCG, bodily coordination features (synchrony, convergence, mimicry, and causality) are positively correlated with PCQ.}
\end{displayquote}

\noindent
For the synchrony attributes, for both IndivPCQ and GroupPCQ we find that the \textit{argmax} and \textit{argmin} variants of lagged correlations are statistically significant attributes ($p=0.003$). This suggests that the time taken to achieve maximum or minimum synchronous coordination has a significant effect on the conversation quality. We also note that for GroupPCQ, only correlation based features from the synchrony category were statistically significant, while other attribute sets (convergence and causality) were found to be statistically insignificant. For IndivPCQ, the \textit{minimum} and \textit{variance} of the convergence attributes were all statistically significant. This suggests that attributes capturing the least converging interacting pairs in a group are relevant to external observers. Moreover, we note that the minimum of the attributes are positively correlated, while the variance are negatively correlated. Further, the \textit{maximum} and \textit{minimum} of the lagged mimicry attributes were also statistically significant attributes. This suggests that pairs with high and low mimicry are relevant for estimating individual experience.

The Joint LASSO results indicate that several other feature sets also have a significant effect on IndivCQ. Along with the \textit{min}, \textit{max}, \textit{argmin}, and \textit{argmax} attributes of the lagged correlation features, the  non-lagged correlation were also significant. Moreover, the post-hoc rank correlation analysis associates different coefficient signs for some of the significant features. For example, lagged mimicry attributes are given negative $\beta$'s by the rank correlation model but positive $\beta$'s by LASSO. This suggests that there exists a non-linear monotonic relationships between these variables and IndivPCQ, causing the LASSO model to fail to explain this relationship, associating them with $\beta\approx0$. One commonality between the two models is that both consider the \textit{lagged} variant of mimicry features to be of more significance that the \textit{lead} variant. For GroupPCQ, the LASSO and rank correlation analysis reveals that when jointly considered with other bodily coordination features, the lagged mimicry and convergence attributes are statistically significant.  

\subsection{Analysis of Feature Extraction and Fusion}

\subsubsection{Influence of Window Sizes}
%%%%%%%%%%%%%%%%%%%%%%%%%%%%%%%%%%%%%%%%%%%%%%%%%%%%%%%
 \begin{figure}[t]
    \begin{subfigure}[]{0.49\columnwidth}
        \centering
        \includegraphics[width=\linewidth]{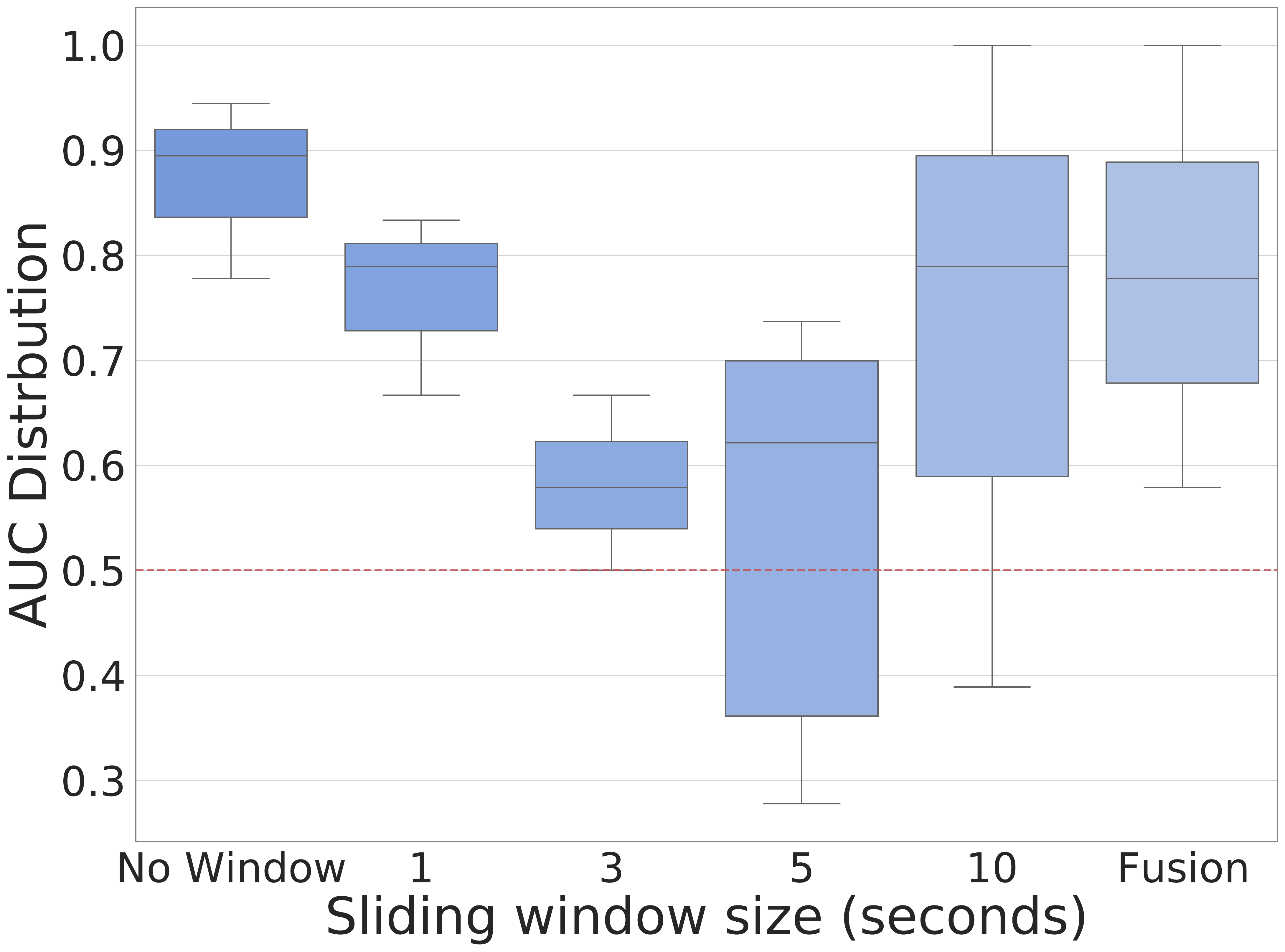}
        \caption{GroupPCQ results}
        \label{fig:pred-window-group}
    \end{subfigure}
    \hfill
    \begin{subfigure}[]{0.49\columnwidth}
        \centering
        \includegraphics[width=\linewidth]{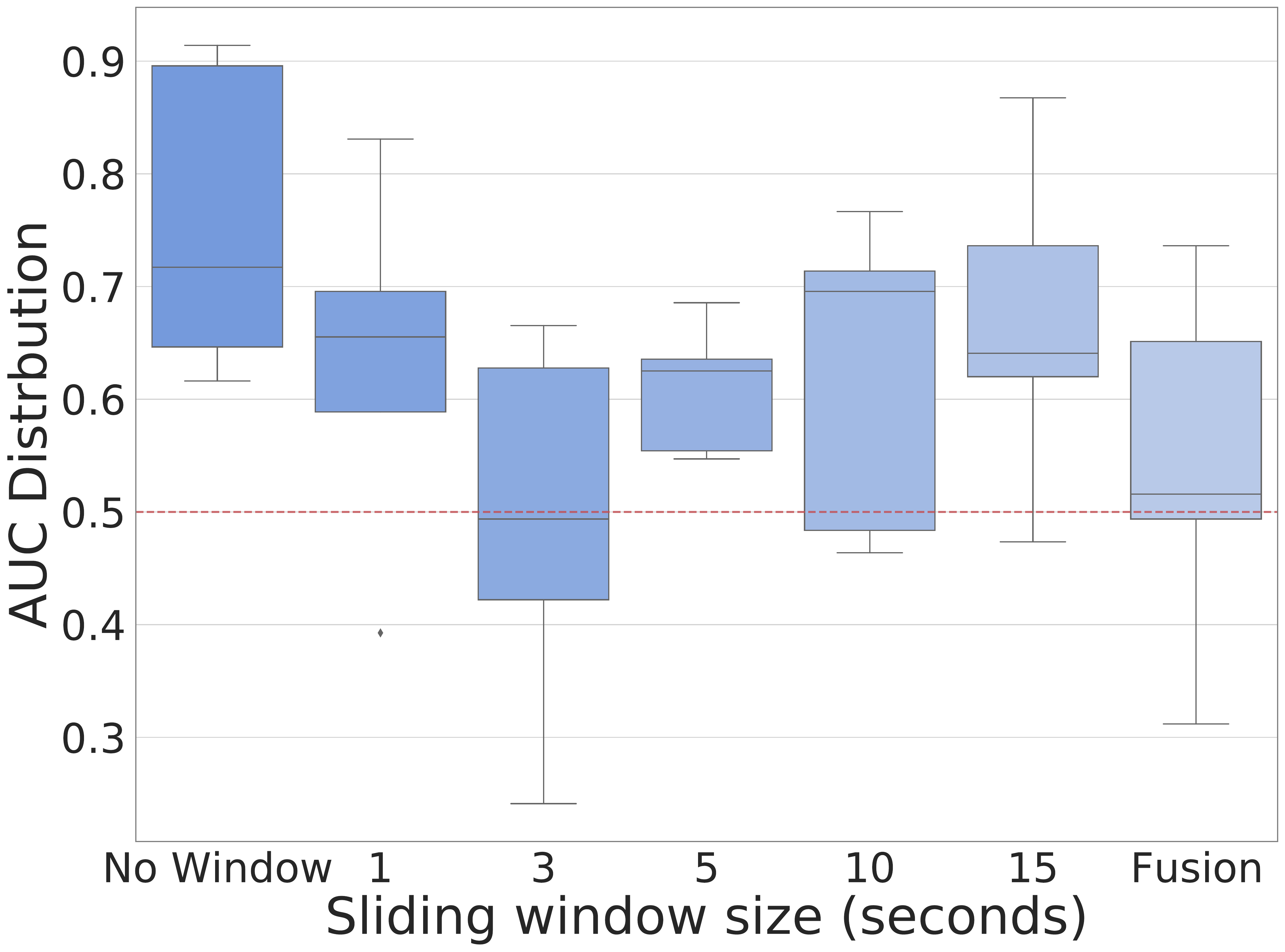}
        \caption{IndivPCQ results}
        \label{fig:pred-window-indiv}
    \end{subfigure}
    \caption{Results of the experiments on the predictive capabilities of different window-sizes.}
    \label{fig:pred-window}
\end{figure}
%%%%%%%%%%%%%%%%%%%%%%%%%%%%%%%%%%%%%%%%%%%%%%%%%%%%%%%
During data preprocessing we extract statistical and spectral features from the accelerometer data using the commonly used sliding window approach \cite{hung2013classifying, gedik2016speaking, kapcak2019estimating}. The choice of the window size influences a trade-off between noise-reduction and information loss. To understand the effect of this choice, we extract features using different window-sizes and evaluate the resulting change in the logistic regression model's predictive capability.

From the results presented in \figurename~\ref{fig:pred-window}, we see that the best performing features are the ones where no sliding-window technique was used, for both GroupPCQ and IndivPCQ. This suggests that the smoothing of accelerometer readings using the sliding window approach results in loss of information, which hurts model performance. The results might also indicate that bodily coordination between interacting pairs occur at finer temporal granularity, which can be captured directly without the sliding-window approach. The model with no sliding-window based features is capable of predicting GroupPCQ with a mean AUC of $0.85 \pm0.07$ and IndivPCQ with a mean AUC of $0.76 \pm0.13$.

\subsubsection{Influence of Fusing Attribute Categories}
In this analysis, we study the influence of fusing different attribute categories on the predictive capability of the logistic regression model.

 %%%%%%%%%%%%%%%%%%%%%%%%%%%%%%%%%%%%%%%%%%%%%%%%%%%%%%%
  \begin{figure}[b]
   \begin{subfigure}[]{0.49\columnwidth}
        \centering
        \includegraphics[width=\linewidth]{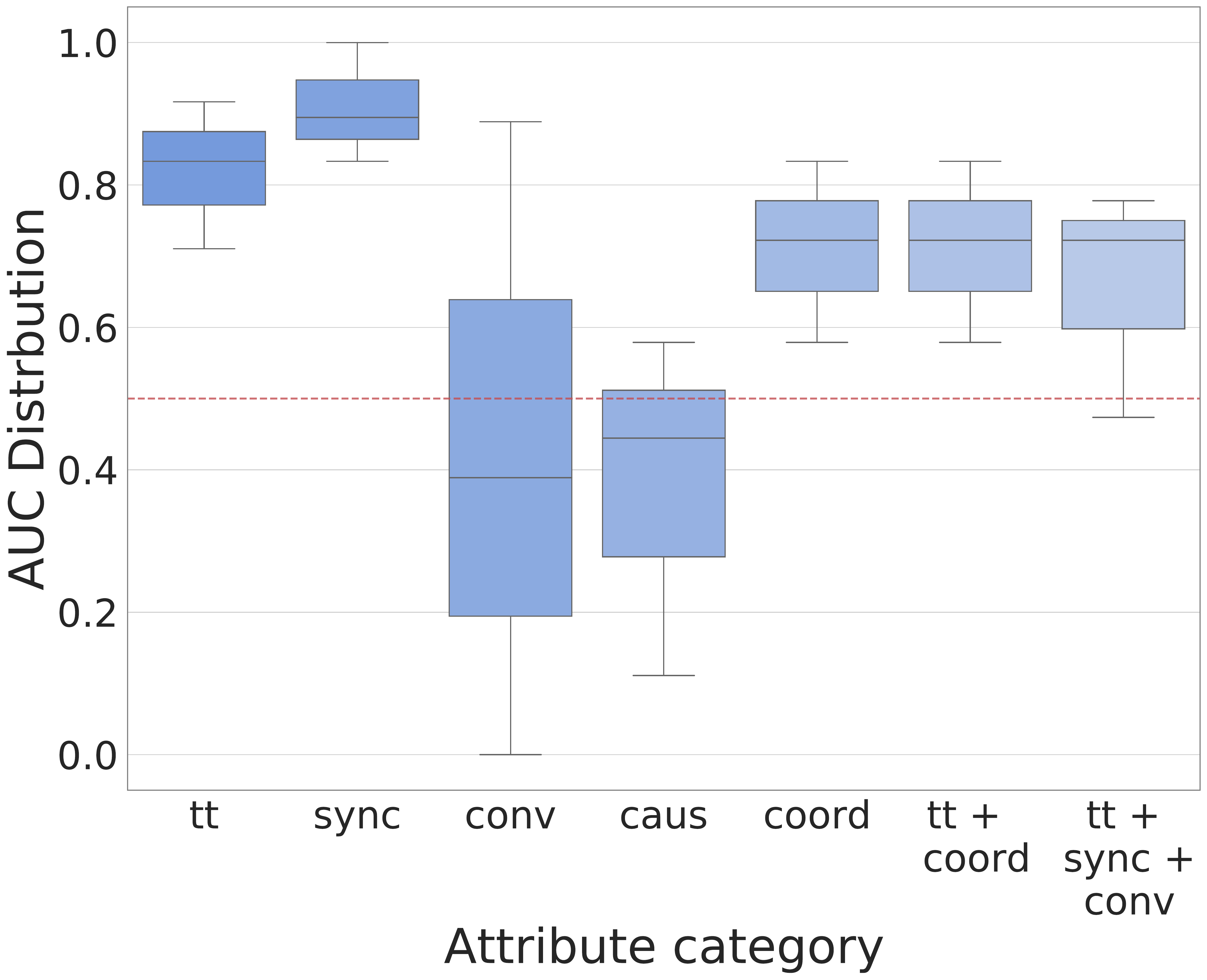}
        \caption{GroupPCQ results}
        \label{fig:pred-feats-group}
    \end{subfigure}
    \hfill
    \begin{subfigure}[]{0.49\columnwidth}
        \centering
        \includegraphics[width=\linewidth]{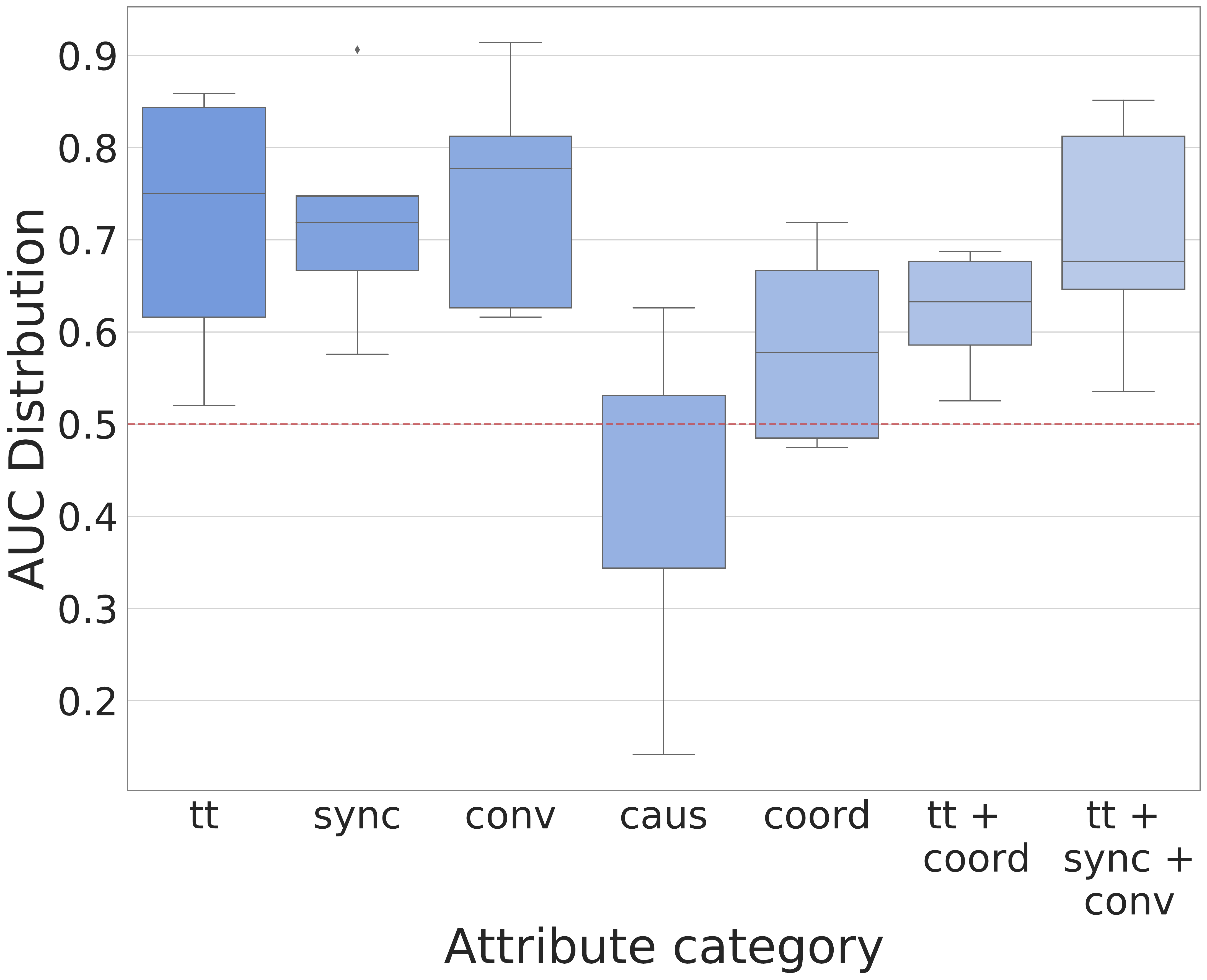}
        \caption{IndivPCQ results}
        \label{fig:pred-feats-indiv}
    \end{subfigure}
    \caption{Predictive performance of different feature fusion approaches. Attribute category labels and indices as in Table~\ref{tab:feats-overview} -- \textit{tt}: Turn-taking (10-12), \textit{sync}: Synchrony (1-4), \textit{caus}: Causality (5-6), \textit{conv}: Convergence (7-9), \textit{coord}: Bodily Coordination (1-9).}
    \label{fig:pred-feats}
\end{figure}
%%%%%%%%%%%%%%%%%%%%%%%%%%%%%%%%%%%%%%%%%%%%%%%%%%%%%%% 

From the GroupPCQ results in \figurename~\ref{fig:pred-feats-group}, we see that the synchrony attributes (mean AUC of $0.89\pm0.04$) and turn-taking attributes (mean AUC of $0.81\pm0.06$), are the best performing attributes. In contrast to the IndivPCQ results in From \figurename~\ref{fig:pred-feats-indiv}, the convergence attributes do not predict GroupPCQ well. Moreover, unlike for IndivPCQ, fusing turn-taking attributes with synchrony and convergence attributes does not improve GroupPCQ prediction, both in-terms of mean and variance AUC. From the IndivPCQ analysis, we see that convergence (mean AUC of $0.75\pm0.12$) and synchrony (mean AUC of $0.72\pm0.12$) based attributes perform well both by themselves and after feature-level fusion (mean AUC of $0.60\pm0.10$). We also observe that although turn-taking attributes are one of the best performing feature sets by themselves (mean AUC of $0.72\pm0.15$), fusing them with bodily coordination attributes reduces the variance of predictions (mean AUC of $0.70\pm0.09$). The results also suggest that synchrony and convergence attributes are best predictors of IndivPCQ, both individually and fused.

\subsubsection{Influence of Feature Aggregators}
The last step of our feature extraction procedure is to use aggregators to combine pairwise features into group-level features, or aggreagate over pairs containing an individual for individual-level modeling in line with approaches of previous works \cite{Nanninga2017, Zhang2018_TeamSense, rapport_muller}. Here we study how different aggregators affect the predictive performance of the logistic regression model.

\begin{figure}[t]
     \begin{subfigure}[]{0.49\columnwidth}
        \centering
        \includegraphics[width=\linewidth]{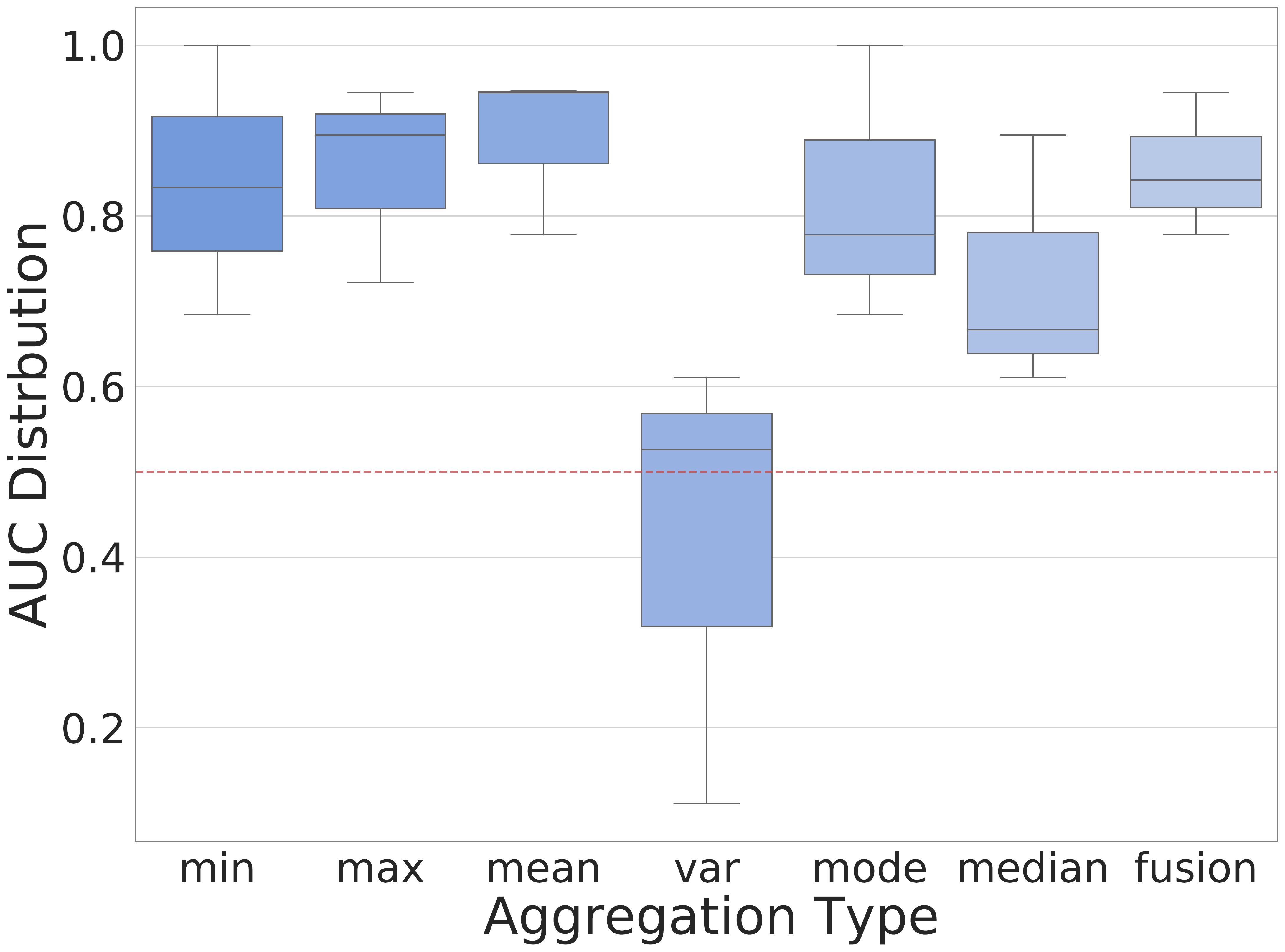}
        \caption{GroupPCQ results}
        \label{fig:pred-groupfeat-group}
    \end{subfigure}
    \hfill
    \begin{subfigure}[]{0.49\columnwidth}
        \centering
        \includegraphics[width=\linewidth]{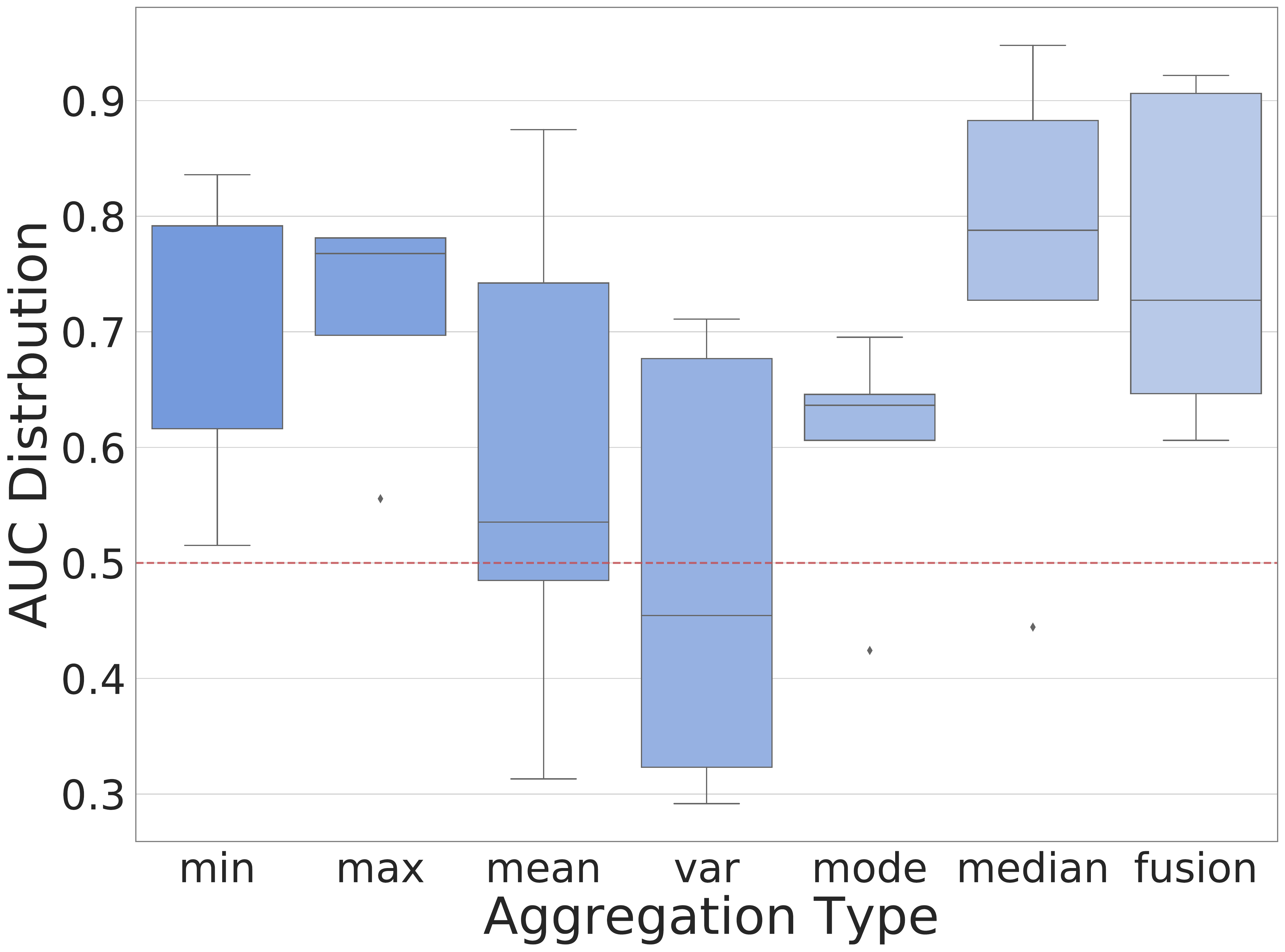}
        \caption{IndivPCQ results}
        \label{fig:pred-groupfeat-indiv}
    \end{subfigure}
    \caption{Predictive performance of feature aggregators.}
    \label{fig:pred-groupfeat}
\end{figure}

From \figurename~\ref{fig:pred-groupfeat}, we see that the the \textit{mean} aggregation of the features performs the best with a mean AUC of $0.89\pm0.08$. The mean is a skewed average. In contrast, for IndivPCQ the unskewed average, the \textit{median}, is the most informative, with an AUC of $0.78\pm0.17$. This is in line with inferences drawn by \citet{Nanninga2017} while studying cohesion in meetings. For both IndivPCQ and GroupPCQ, the \textit{variance} aggregator performs worst.

 %%%%%%%%%%%%%%%%%%%%%%%%%%%%%%%%%%%%%%%%%%%%%%%%%%%%%%%
%%%%%%%%%%%%%%%%%%%%%%%%%%%%%%%%%%%%%%%%%%%%%%%%%%%%%%%

\section{Conclusion} \label{sec:conclusion}

In this work, we have conceptualized, validated, and analyzed a perceived measure of conversation quality by unifying overlapping constructs thus far largely studied in isolation. 
While our core motivation has been to gain insight into how people perceive the individual and group experiences of others, we do not claim that our proposed method measures any `one true experience'. On the contrary, we suggest that these perceptions are indicative of empathized gestalt impressions people draw of others' experience as it unfolds. We argue that such a perceived measure should complement other self-reported measures of experience to gain richer insight into how these differ, and identify the contextual factors that influence the perceptions. In the same vein, it is important to note that the data analysed here was from spontaneous interactions in a single setting, that of mingling interactions following a speed-dating event. So our findings pertaining to individual features being indicative of PCQ ought to be interpreted within the scope of such a social context rather than being indicative of social behavior in all spontaneous interactions. As dedicated techniques for recording in-the-wild spontaneous interactions non-invasively \cite{raman2020modular} continue to advance, it would be interesting to compare the effects of different social settings on the perception of PCQ using our proposed instrument.

Our operationalization of a conversing group follows the widely used framework F-formation \cite{kendon1990conducting}. However, recent evidence suggests that there might be multiple simultaneous conversations within a single F-formation for groups with more than four participants \cite{Raman2019}. It would therefore be interesting for future work to study PCQ within a single conversation floor rather than for the whole F-formation.

\ifCLASSOPTIONcompsoc
  \section*{Acknowledgments}
\else
  \section*{Acknowledgment}
\fi
This research was partially funded by the Netherlands Organization for Scientific Research (NWO) under the MINGLE project number 639.022.606. We thank Swathi Yogesh, Divya Suresh Babu, and Nakul Ramachandran for their time and patience in annotating the dataset.

\ifCLASSOPTIONcaptionsoff
  \newpage
\fi



%\bibliographystyle{IEEEtran}

% H.~Kopka and P.~W. Daly, \emph{A Guide to {\LaTeX}}, 3rd~ed.\hskip 1em plus
%   0.5em minus 0.5em\relax Harlow, England: Addison-Wesley, 1999.

% \end{thebibliography}

\bibliographystyle{IEEEtranN}
\bibliography{conv_biblio}


% \end{IEEEbiography}

% BIO Chirag
\begin{IEEEbiography}[{\includegraphics[width=1in,height=1.25in,clip,keepaspectratio]{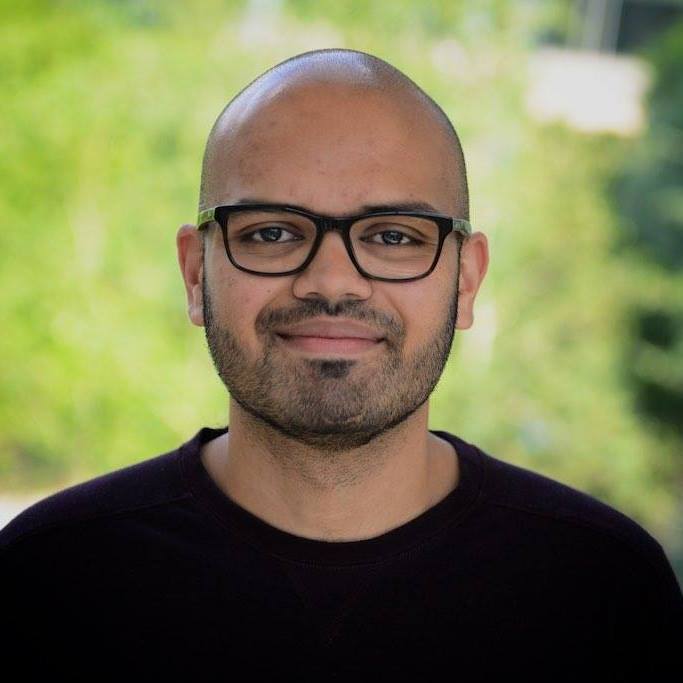}}] {Chirag Raman}
received his B.Eng in Information Technology from the University of Mumbai, India in 2010 and his Masters in Entertainment Technology from Carnegie Mellon University, USA in 2013. Between 2013 and 2018 he worked as a Research Associate at Disney Research, a Lead iOS and UX developer at ProductionPro, and a Senior Research Engineer at the Language Technologies Institute at Carnegie Mellon University. Since 2018 he has been pursuing his Ph.D. as part of the Socially Perceptive Computing and Pattern Recognition Labs at TU Delft, The Netherlands. His research interests include multimodal machine learning, generative modeling, affective computing, computer vision, and computer graphics.
\end{IEEEbiography}

% BIO Navin
\begin{IEEEbiography}[{\includegraphics[width=1in,height=1.25in,clip,keepaspectratio]{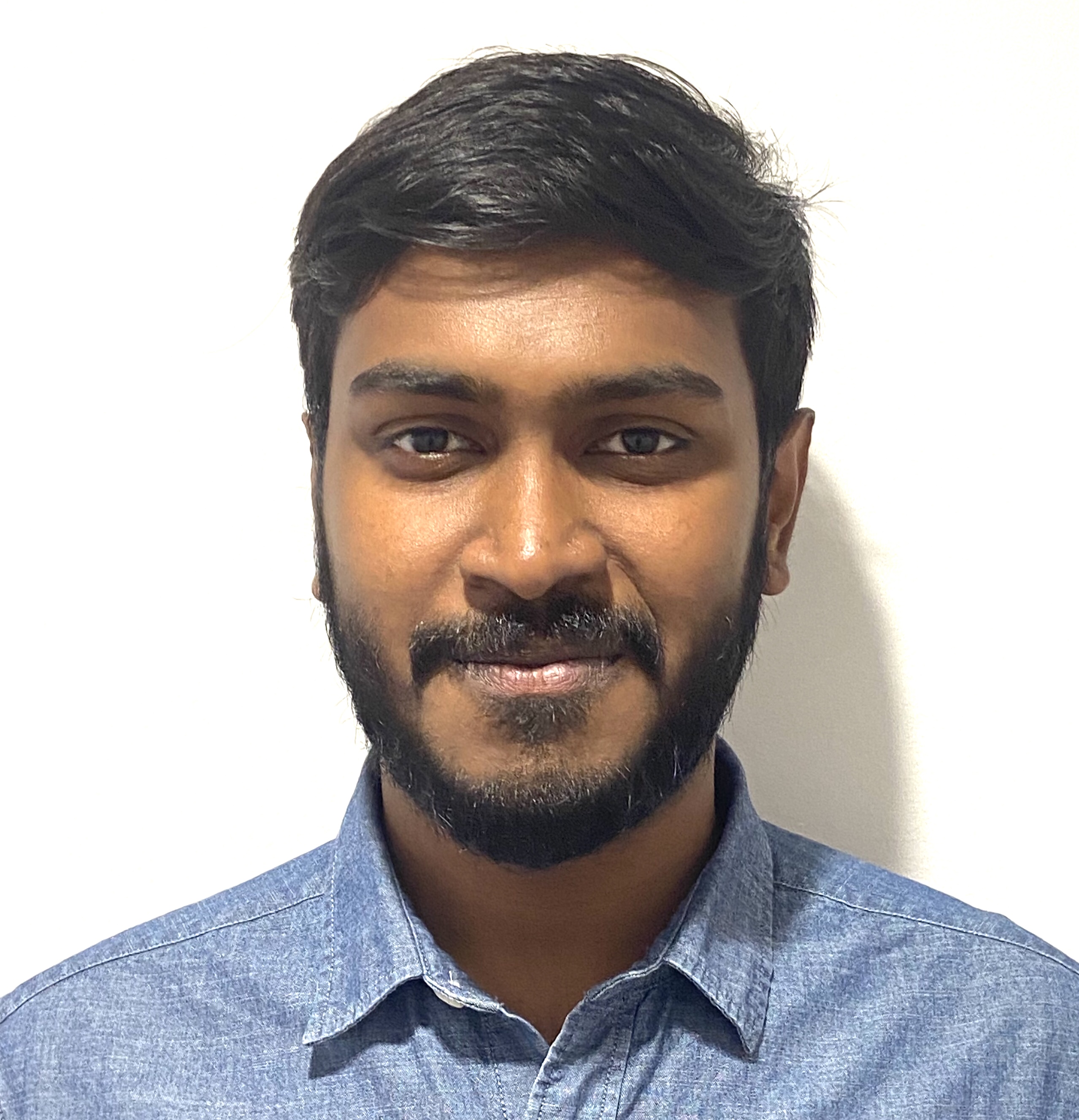}}]{Navin Raj Prabhu}
received the B.Tech degree in Computer Science from SRM University, India, in 2015, and the MS degree in Computer Science from the Intelligent Systems Department at the Delft University of Technology, Delft, The Netherlands, in 2020. Currently, he is a PhD student at the Signal Processing Lab and Organisation Psychology Lab, University of Hamburg, Hamburg, Germany. His research interests include affective computing, social signal processing, deep learning, uncertainty modelling, speech signal processing, and group affect. 
\end{IEEEbiography}

% BIO Hayley
\begin{IEEEbiography}[{\includegraphics[width=1in,height=1.25in,clip,keepaspectratio]{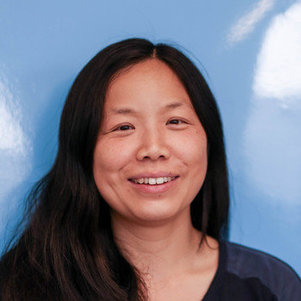}}]{Hayley Hung}
is an Associate Professor in the Socially Perceptive Computing Lab at TU Delft, The Netherlands, where she works since 2013. Between 2010-2013 she held a Marie Curie Fellowship at the Intelligent Systems Lab at the University of Amsterdam. Between 2007-2010 she was a post-doctoral researcher at IDIAP Research Institute in Switzerland. She obtained her PhD in Computer Vision from Queen Mary University of London in 2007. Her research interests are social computing, social signal processing, computer vision, and machine learning.
\end{IEEEbiography}


% % insert where needed to balance the two columns on the last page with
% %\newpage



%\vfill



\includepdf[pages=-]{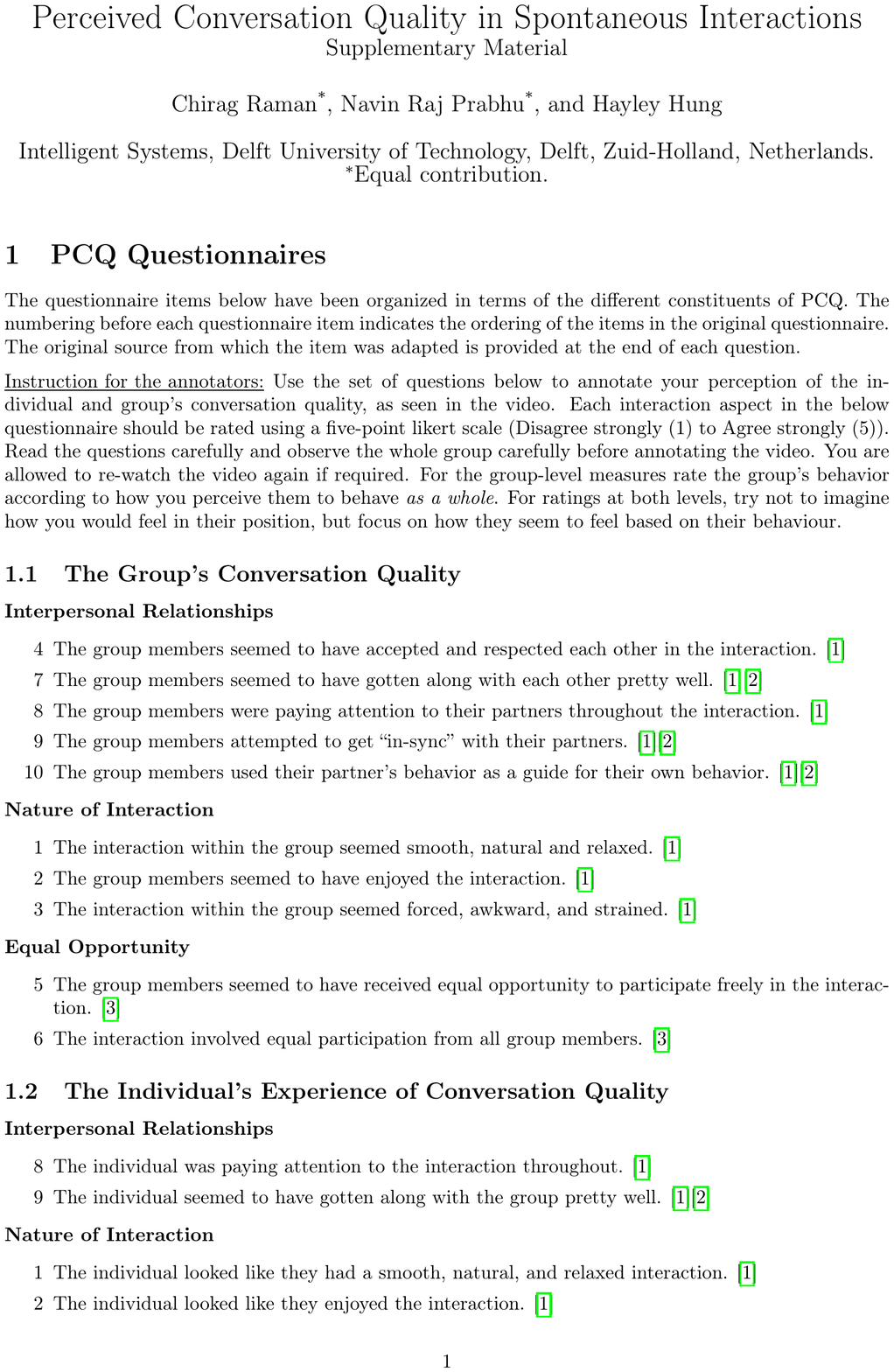}

\end{document}

\maketitle

%%%%%%%%%%%%%%%%%%%%%%%%%%%%%%%%%%%%%%%%%%%%%%%%%%%%%%%%%%%%%%%%%%%%
\section{PCQ Questionnaires}

The questionnaire items below have been organized in terms of the different constituents of PCQ. The numbering before each questionnaire item indicates the ordering of the items in the original questionnaire. The original source from which the item was adapted is provided at the end of each question.

\medskip
\noindent\underline{Instruction for the annotators:} Use the set of questions below to annotate your perception of the individual and group’s conversation quality, as seen in the video. Each interaction aspect in the below questionnaire should be rated using a five-point likert scale (Disagree strongly (1) to Agree strongly (5)). Read the questions carefully and observe the whole group carefully before annotating the video. You are allowed to re-watch the video again if required. For the group-level measures rate the group's behavior according to how you perceive them to behave \textit{as a whole}. For ratings at both levels, try not to imagine how you would feel in their position, but focus on how they seem to feel based on their behaviour.

\subsection{The Group’s Conversation Quality} \label{questionnaire-group}
\textbf{Interpersonal Relationships}
\begin{enumerate}
    \item[4] The group members seemed to have accepted and respected each other in the interaction. \cite{Cuperman2009}
    \item[7] The group members seemed to have gotten along with each other pretty well. \cite{Cuperman2009}\cite{Jaques2016}
    \item[8] The group members were paying attention to their partners throughout the interaction. \cite{Cuperman2009}
    \item[9] The group members attempted to get ``in-sync'' with their partners. \cite{Cuperman2009}\cite{Jaques2016}
    \item[10] The group members used their partner's behavior as a guide for their own behavior. \cite{Cuperman2009}\cite{Jaques2016}
\end{enumerate}
\textbf{Nature of Interaction}
\begin{enumerate}
    \item[1] The interaction within the group seemed smooth, natural and relaxed. \cite{Cuperman2009}
    \item[2] The group members seemed to have enjoyed the interaction. \cite{Cuperman2009}
    \item[3] The interaction within the group seemed forced, awkward, and strained. \cite{Cuperman2009}
\end{enumerate}
\textbf{Equal Opportunity}
\begin{enumerate}
    \item[5] The group members seemed to have received equal opportunity to participate freely in the interaction. \cite{Lindley2013}
    \item[6] The interaction involved equal participation from all group members. \cite{Lindley2013}
\end{enumerate}

\subsection{The Individual’s Experience of Conversation Quality}\label{questionnaire-indiv}
\textbf{Interpersonal Relationships}
\begin{enumerate}
    \item[8] The individual was paying attention to the interaction throughout. \cite{Cuperman2009}
    \item[9] The individual seemed to have gotten along with the group pretty well. \cite{Cuperman2009}\cite{Jaques2016}
\end{enumerate}
\textbf{Nature of Interaction}
\begin{enumerate}
    \item[1] The individual looked like they had a smooth, natural, and relaxed interaction. \cite{Cuperman2009}
    \item[2] The individual looked like they enjoyed the interaction. \cite{Cuperman2009}
    \item[3] The individual’s interaction seemed to be forced, awkward, and strained. \cite{Cuperman2009}
    \item[4] The individual looked like they had a pleasant and an interesting interaction. \cite{Cuperman2009}
    \item[5] The individual appeared uncomfortable during the interaction. \cite{Cuperman2009}
    \item[10] The individual appeared self-conscious during the interaction. \cite{Cuperman2009}
\end{enumerate}
\textbf{Equal Opportunity}
\begin{enumerate}
    \item[6] The individual attempted to take the lead in the conversation. \cite{Jaques2016}\cite{Cerekovic2014}
    \item[7] The individual looked like they experienced a free-for-all interaction. \cite{Lindley2013}
\end{enumerate}

%%%%%%%%%%%%%%%%%%%%%%%%%%%%%%%%%%%%%%%%%%%%%%%%%%%%%%%%%%%%%%%%%%%%
\FloatBarrier
\section{Additional Figures}

\begin{figure}[h]
    % Group _ Level plots
    \begin{subfigure}[b]{0.48\linewidth}    
        \centering
        \includegraphics[width=\linewidth]{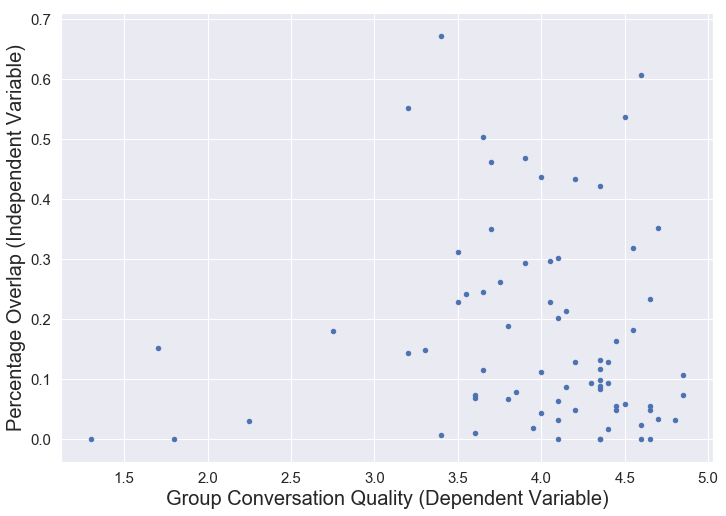}
        % \captionsetup{justification=centering, margin=0.5cm}
        \caption{Scatter plot revealing the relationship between Percentage of Overlap (Independent Variable) and the Group Conversation Quality (Dependent Variable).}
        \label{fig:overlapVSConvQ-group}
    \end{subfigure}%
    \hfill
    \begin{subfigure}[b]{0.48\linewidth}
        \centering
        \includegraphics[width=\linewidth]{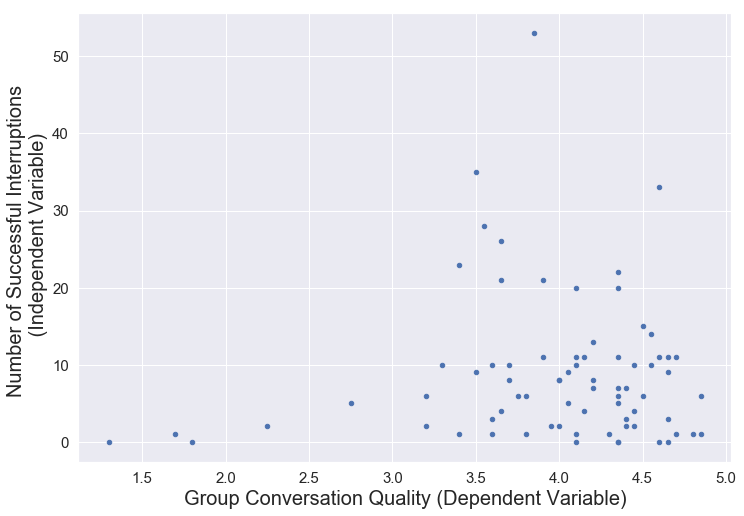}
        % \captionsetup{justification=centering, margin=0.5cm}
        \caption{Scatter plot revealing the relationship between Number of Successful Interruptions (Independent Variable) and the Group Conversation Quality (Dependent Variable).}
        \label{fig:interupVSConvQ-group}
    \end{subfigure}%'
    % \captionsetup{justification=centering, margin=0.5cm}
    \vfill
    \medskip
    % Indiv _ Level plots
    \begin{subfigure}[b]{0.48\linewidth}    
        \centering
        \includegraphics[width=\linewidth]{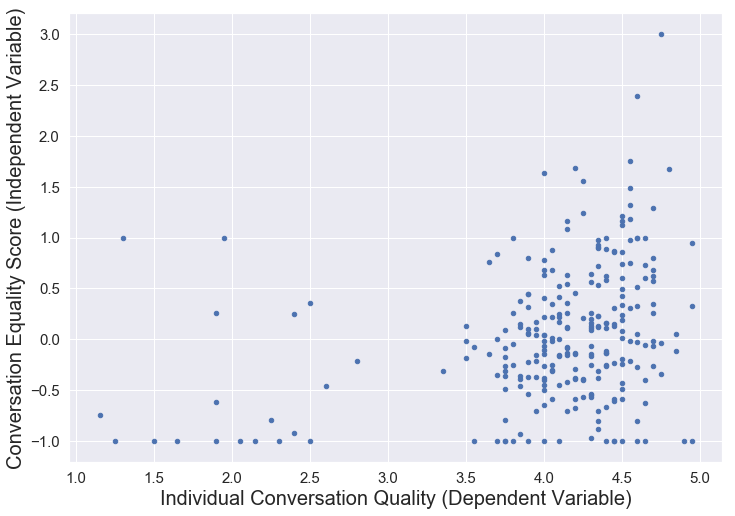}
        % \captionsetup{justification=centering, margin=0.5cm}
        \caption{Scatter plot revealing the relationship between Conversation Equality (Independent Variable) and the Individual Conversation Quality (Dependent Variable).}
        \label{fig:eqVSconvq-indiv}
    \end{subfigure}%
    \hfill
    \begin{subfigure}[b]{0.48\linewidth}
        \centering
        \includegraphics[width=\linewidth]{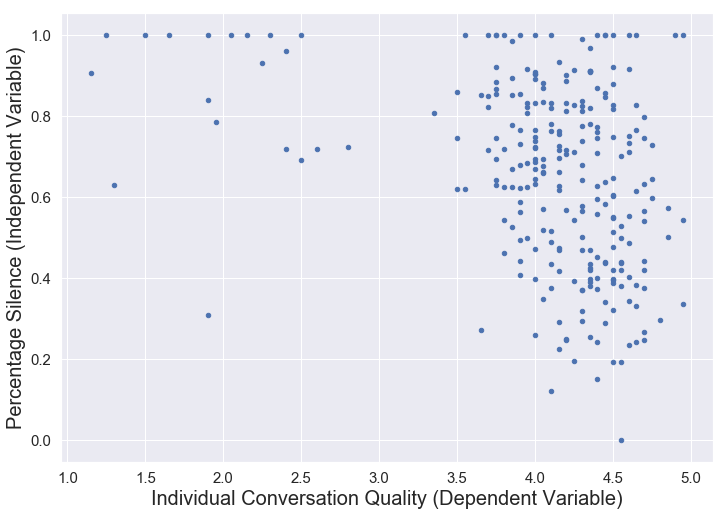}
        % \captionsetup{justification=centering, margin=0.5cm}
        \caption{Scatter plot revealing the relationship between Percentage of Silence (Independent Variable) and the Individual Conversation Quality (Dependent Variable).}
        \label{fig:silenceVSconvq-indiv}
    \end{subfigure}%'
    \medskip
    % \captionsetup{justification=centering, margin=0.5cm}
    \caption{Scatter plots with respect to few Independent Variables and the Dependent Variables of Conversation Quality. The scatter plots are a qualitative analysis of the independent variable's variance ($\sigma^2$) conditioned to the dependent variable of Conversation Quality, and thus examine the Exogeneity and Homoscedasticity of the dataset.}
    \label{fig:rationale-qls}
\end{figure}

%%%%%%%%%%%%%%%%%%%%%%%%%%%%%%%%%%%%%%%%%%%%%%%%%%%%%%%%%%%%%%%%%%%%
% \FloatBarrier
\newpage
\printbibliography